\RequirePackage[loading]{tracefnt}
\documentclass{article}
\usepackage[
            CJKbookmarks=true,
            bookmarksnumbered=true,
            bookmarksopen=true,
            colorlinks=true,
                        citecolor=magenta,
            linkcolor=blue,
            anchorcolor=red,
            urlcolor=blue
            ]{hyperref}

\usepackage{tabu}
\usepackage[utf8]{inputenc} 
\usepackage[T1]{fontenc}    
\usepackage{url}            
\usepackage{booktabs}       
\usepackage{amsfonts}       
\usepackage{nicefrac}       
\usepackage{microtype}      

\usepackage{xcolor}
\definecolor{MyGreen1}{RGB}{20,180,40}
\usepackage{multirow,caption,tikz,float}
\usepackage[linesnumbered,ruled,vlined, algo2e, resetcount]{algorithm2e}

\usepackage{times,nicefrac,color,mathrsfs,float}
\usepackage{multirow,caption,tikz}
\usepackage{subcaption}
\usepackage{comment}

\usetikzlibrary{shapes.misc, positioning}
\usetikzlibrary{decorations.pathreplacing}

\tikzset{
  block/.style    = {draw, thick, rectangle, minimum width = 2em},
sblock/.style      = {draw, thick, rectangle, minimum height = 2em,
minimum width = 2em}, 
}

\usepackage{color,fancyhdr}
\usepackage{amsmath,amssymb,amsthm,amsfonts,dsfont}
\usepackage{enumerate} 

\usepackage{algorithmic, algorithm}
\usepackage{mathtools}

\newcommand{\Expect}{\mathbb{E}}

\newcommand{\norm}[1]{\|#1\|}

\renewcommand{\tilde}{\widetilde}

\newcommand{\RM}{\text{RM}}
\newcommand{\KO}{\text{KO}}

\usepackage{comment}

\usepackage{thmtools}

\usepackage{prettyref,xspace}
\usepackage{tikz}

\newrefformat{cond}{Condition~\ref{#1}}
\newrefformat{eq}{(\ref{#1})}
\newrefformat{thm}{Theorem~\ref{#1}}
\newrefformat{th}{Theorem~\ref{#1}}
\newrefformat{chap}{Chapter~\ref{#1}}
\newrefformat{sec}{Section~\ref{#1}}
\newrefformat{algo}{Algorithm~\ref{#1}}
\newrefformat{fig}{Figure~\ref{#1}}
\newrefformat{tab}{Table~\ref{#1}}
\newrefformat{rmk}{Remark~\ref{#1}}
\newrefformat{clm}{Claim~\ref{#1}}
\newrefformat{def}{Definition~\ref{#1}}
\newrefformat{cor}{Corollary~\ref{#1}}
\newrefformat{lmm}{Lemma~\ref{#1}}
\newrefformat{prop}{Proposition~\ref{#1}}
\newrefformat{pr}{Proposition~\ref{#1}}
\newrefformat{app}{Appendix~\ref{#1}}
\newrefformat{prob}{Problem~\ref{#1}}
\newrefformat{ques}{Question~\ref{#1}}
\newrefformat{note}{Note~\ref{#1}}
\newrefformat{assump}{Assumption~\ref{#1}}
\newrefformat{issue}{Issue ~\ref{#1}}
\newrefformat{fix}{Fix ~\ref{#1}}

\newcommand{\reals}{\mathbb{R}}

\newcommand{\prob}[1]{\mathbb{P}\left[#1\right]}

\newcommand{\calL}{\mathcal{L}}

\newcommand{\vect}[1]{\boldsymbol{#1}}

\newcommand{\bc}{\vect{c}}

\newcommand{\bma}{\vect{m}}

\newcommand{\bu}{\vect{u}}
\newcommand{\bv}{\vect{v}}

\newcommand{\bx}{\vect{x}}
\newcommand{\by}{\vect{y}}

\newcommand{\bG}{\vect{G}}

\newcommand{\bL}{\vect{L}}

\newcommand{\define}{\triangleq}

\newcommand{\ie}{i.e.\xspace}

\mathchardef\mhyphen="2D

\definecolor{MyGreen1}{RGB}{20,180,40}

\usepackage{siunitx}

\usepackage[accepted]{icml2021}

\icmltitlerunning{KO codes}

\begin{document}

\twocolumn[
\icmltitle{KO codes: Inventing Nonlinear Encoding and Decoding for Reliable Wireless Communication via Deep-learning}



\icmlsetsymbol{equal}{*}

\begin{icmlauthorlist}
\icmlauthor{Ashok Vardhan Makkuva}{equal,uiuc}
\icmlauthor{Xiyang Liu}{equal,uw}
\icmlauthor{Mohammad Vahid Jamali}{um}
\icmlauthor{Hessam Mahdavifar}{um}
\icmlauthor{Sewoong Oh}{uw}
\icmlauthor{Pramod Viswanath}{uiuc}
\end{icmlauthorlist}

\icmlaffiliation{uiuc}{Department of Electrical and Computer Engineering, University of Illinois at Urbana-Champaign}
\icmlaffiliation{um}{Department of Electrical Engineerign and Computer Science, University of Michigan} 
\icmlaffiliation{uw}{Paul G. Allen School of Computer Science \& Engineering, University of Washington}

\icmlcorrespondingauthor{Ashok, Xiyang}{makkuva2@illinois.edu, xiyangl@cs.washington.edu}

\icmlkeywords{communication, coding theory, information theory}

\vskip 0.3in
]



\printAffiliationsAndNotice{\icmlEqualContribution}  

\newcommand{\Ashok}[1]{{\color{magenta}[Ashok: #1]}}
\newcommand{\Xiyang}[1]{{\color{blue}[Xiyang: #1]}}
\newcommand{\Sewoong}[1]{{\color{red}[Sewoong: #1]}}
\newcommand{\pv}[1]{{\color{red}[Pramod: #1]}}


\begin{abstract}
Landmark codes underpin reliable physical layer communication, e.g.,  Reed-Muller, BCH, Convolution, Turbo, LDPC  and Polar codes: each is a linear code and represents a mathematical breakthrough. The impact on humanity is huge: each of these codes has been used in global wireless communication standards (satellite, WiFi, cellular). Reliability of communication over the classical additive white Gaussian noise (AWGN) channel enables benchmarking and ranking of the different codes. In this paper, we construct KO codes, a computationaly efficient family of deep-learning driven (encoder, decoder) pairs that outperform the state-of-the-art reliability performance on the standardized AWGN channel. KO codes  beat state-of-the-art Reed-Muller and Polar codes, under the low-complexity successive cancellation decoding, in the challenging short-to-medium block length regime on the AWGN channel.  We show that the gains of KO codes are primarily due to the nonlinear mapping of information bits directly to transmit real symbols (bypassing modulation) and yet possess  an efficient, high performance   decoder.  The key technical innovation that renders this possible is design of a novel family of neural architectures inspired by the computation tree of the {\bf K}ronecker {\bf O}peration (KO) central to Reed-Muller and Polar codes. These architectures pave way for the discovery of a much richer class of hitherto unexplored nonlinear algebraic structures. The code is available at   \href{https://github.com/deepcomm/KOcodes}{https://github.com/deepcomm/KOcodes}.

\end{abstract} 


\section{Introduction}
\label{sec:intro}



Physical layer communication underpins the information age (WiFi, cellular, cable and satellite modems). 
Codes, composed of encoder and decoder pairs, are the basic mathematical objects enabling reliable communication: encoder  maps original data bits into a longer sequence, and decoders  map the received sequence  
to the original bits. Reliability is precisely measured: bit  error rate (BER) measures the fraction of input bits that were incorrectly decoded; block error rate (BLER) measures the fraction of times at least one of the original data bits was incorrectly decoded. 

Landmark codes include Reed-Muller (RM), BCH, Turbo, LDPC  and Polar codes \citep{richardson2008modern}: each is a linear code 
and represents a mathematical breakthrough discovered over a span of six decades. The impact on humanity is huge: each of these codes has been used in global communication standards over the past six decades. These codes essentially operate at the information-theoretic limits of reliability over the  additive white Gaussian noise (AWGN) channel, when the number of information bits is large, the so-called ``large block length" regime. In the small and medium block length regimes, the state-of-the-art codes are {\em algebraic}: encoders and decoders are invented based on specific linear algebraic constructions over the binary and higher order fields and rings. Especially prominent binary algebraic codes are RM codes and closely related polar codes, whose encoders are recursively defined as Kronecker products of a simple linear operator and constitute the state of the art in small-to-medium block length regimes. 

Inventing new codes  is a major intellectual activity 
both in academia and the wireless industry; this is driven by emerging practical applications, e.g., low block length regime in  Internet of Things  \citep{ma2019high}. 
 The core challenge is that the space of codes is very vast and the sizes astronomical; for instance a rate $1/2$ code over even $100$ information bits involves designing $2^{100}$ codewords in a $200$ dimensional space. Computationally efficient encoding and decoding procedures are a must, apart from high reliability. 
 Thus, although a random code is information theoretically optimal, neither encoding nor decoding is computationally efficient. The mathematical landscape of computationally efficient codes has been plumbed over the decades by some of the finest mathematical minds, resulting in two distinct families of codes: \textit{algebraic codes} (RM, BCH -- focused on properties of polynomials) 
 and \textit{graph codes} (Turbo, LDPC -- based on sparse graphs and statistical physics). The former is deterministic and involves discrete mathematics, while the latter harnesses randomness, graphs, and statistical physics to behave like a pseudorandom code. A major open question is the invention of  new codes, and especially fascinating would be a family of codes outside of these two classes. 

Our major result is the invention of a new family of codes, called \KO~codes, that have features of both code families: they are nonlinear generalizations of the Kronecker operation underlying the algebraic codes (e.g., Reed-Muller) parameterized by neural networks; the parameters are learnt in an end-to-end training paradigm in a data driven manner. Deep learning (DL) has transformed several domains of human endeavor that have traditionally relied heavily on  mathematical ingenuity,  e.g., game playing (AlphaZero \cite{silver2018general}), biology (AlphaFold \cite{senior2019protein}), and physics (new laws \cite{udrescu2020ai}). Our results can be viewed as an added domain to the successes of DL in inventing mathematical structures. 

A linear encoder is defined by a {\em generator matrix}, which maps information bits to a codeword. 
The RM and the Polar families construct their generator matrices by recursively applying the Kronecker product operation to a simple two-by-two matrix and then selecting rows from the resulting matrix. The careful choice in selecting these rows is driven by the desired algebraic structure of the code, which is central to achieving the large {\em minimum}  pairwise distance between two codewords, a hallmark of the algebraic family. This  encoder can be alternatively represented by a computation graph. The recursive Kronecker product corresponds to a complete binary tree, and row-selection corresponds to freezing a set of leaves in the tree, 
which we refer to as a ``Plotkin  tree", inspired by the pioneering construction in %
\cite{plotkin60}. 

The Plotkin tree skeleton allows us to tailor  a new neural network architecture: 
we expand the algebraic family of codes by replacing the (linear) Plotkin construction with a non-linear operation parametrized by neural networks. 
The parameters are discovered by training the encoder with a matching decoder, that has the matching Plotkin tree as a skeleton, to minimize the error rate over (the unlimited) samples generated on AWGN channels.

Algebraic and the original RM codes promise a large worst-case pairwise distance \cite{alon2005testing}. This ensures that RM codes achieve capacity in the large block length limit \cite{kudekar2017reed}. However, for short block lengths, they are too conservative as we are interested in the average-case reliability. This is the gap \KO~codes  exploit: we seek a better average-case reliability and not the minimum pairwise distance. 

\vspace{-0.2cm}
\begin{figure}[h]
    \centering
\includegraphics[width=.4\textwidth]{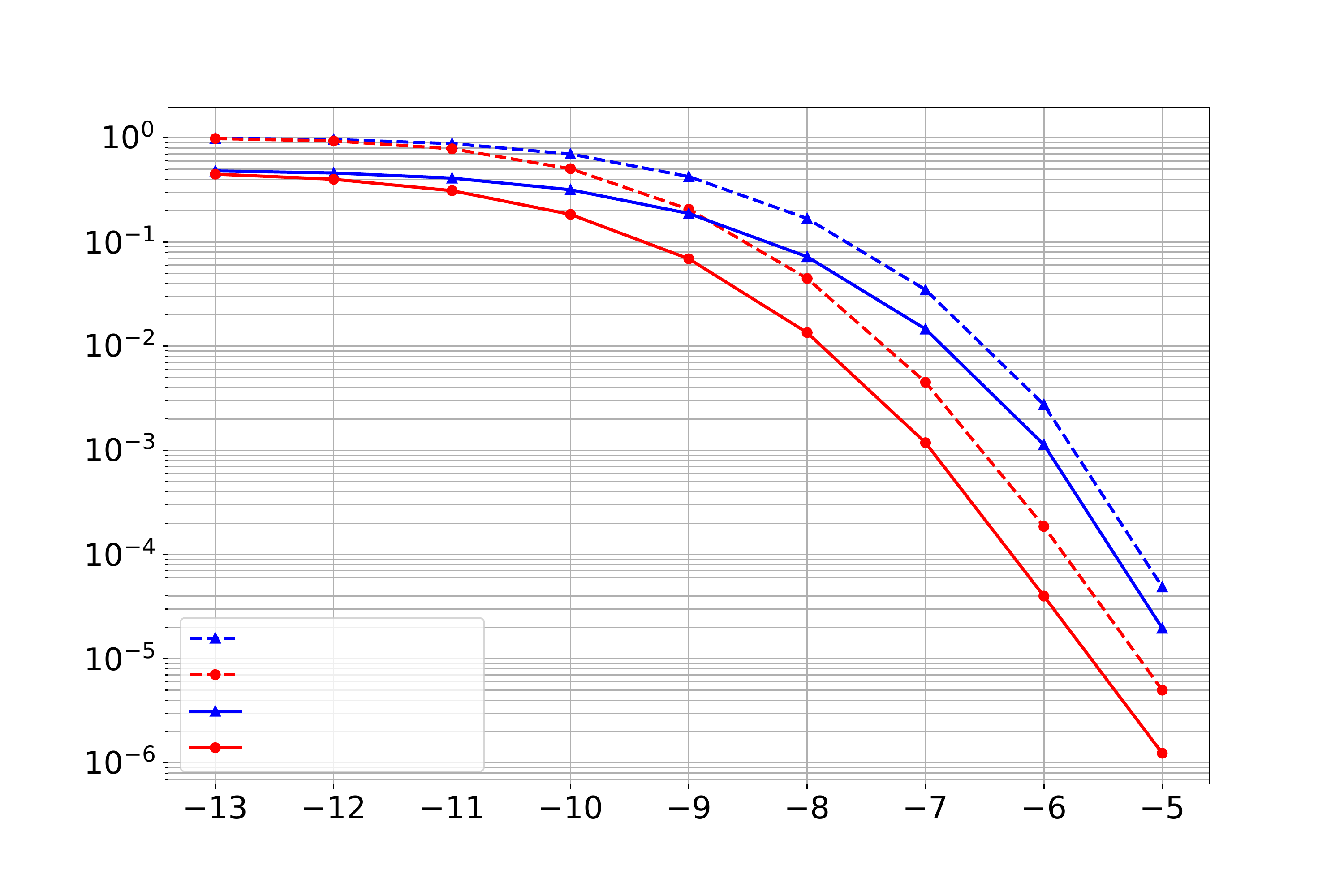}
    \put(-157,35){\fontsize{4}{5}\selectfont RM(9,2) BLER}
    \put(-157,30){\fontsize{4}{5}\selectfont \KO(9,2) BLER}
    \put(-157,25){\fontsize{4}{5}\selectfont RM(9,2) BER}
    \put(-157,20){\fontsize{4}{5}\selectfont \KO(9,2) BER}
    \put(-153,0){\footnotesize Signal-to-noise ratio (SNR) [dB]}
    \put(-192,63){\rotatebox[origin=t]{90}{\footnotesize Error rate}}
    \caption{ $\KO(9,2)$, discovered by training a neural network with a carefully chosen architecture in \S\ref{sec:neural_ko}, significantly improves upon state-of-the-art RM$(9,2)$ both in BER and BLER. (For both codes, the code block length is $2^9 = 512$ and the number of transmitted message bits is ${9 \choose 0}+{9 \choose 1}+{9 \choose 2} = 55$. Also, both codes are decoded using successive cancellation decoding with similar decoding complexity)}
    \label{fig:1} 
\end{figure}
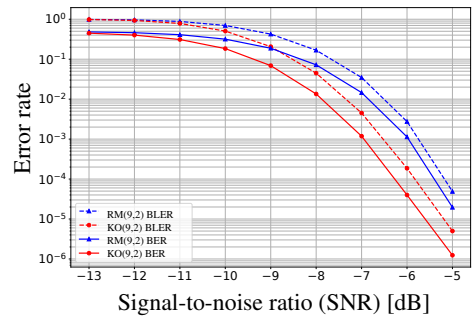

Figure~\ref{fig:1} illustrates the gain for the example of RM$(9,2)$ code. Using the Plotkin tree of $\RM(9,2)$ code as a skeleton, we design the $\KO(9,2)$ code architecture and train on samples simulated over an AWGN channel. We discover a novel non-linear code and a corresponding efficient decoder that improves significantly over the $\RM(9,2)$ code baseline, assuming both codes are decoded using successive cancellation decoding with similar decoding complexity. Analyzing the pairwise distances between two codewords reveals a surprising fact. The histogram for \KO~code nearly matches that of a random Gaussian codebook. The skeleton of the architecture from an algebraic family of codes, the training process with a variation of the stochastic gradient descent, and the simulated AWGN channel have worked together to discover a novel family of codes that harness the benefits of both algebraic and pseudorandom constructions. 

\vspace{-0.18in}
\begin{figure}[H] 
    \centering
    \includegraphics[width=.43\textwidth]{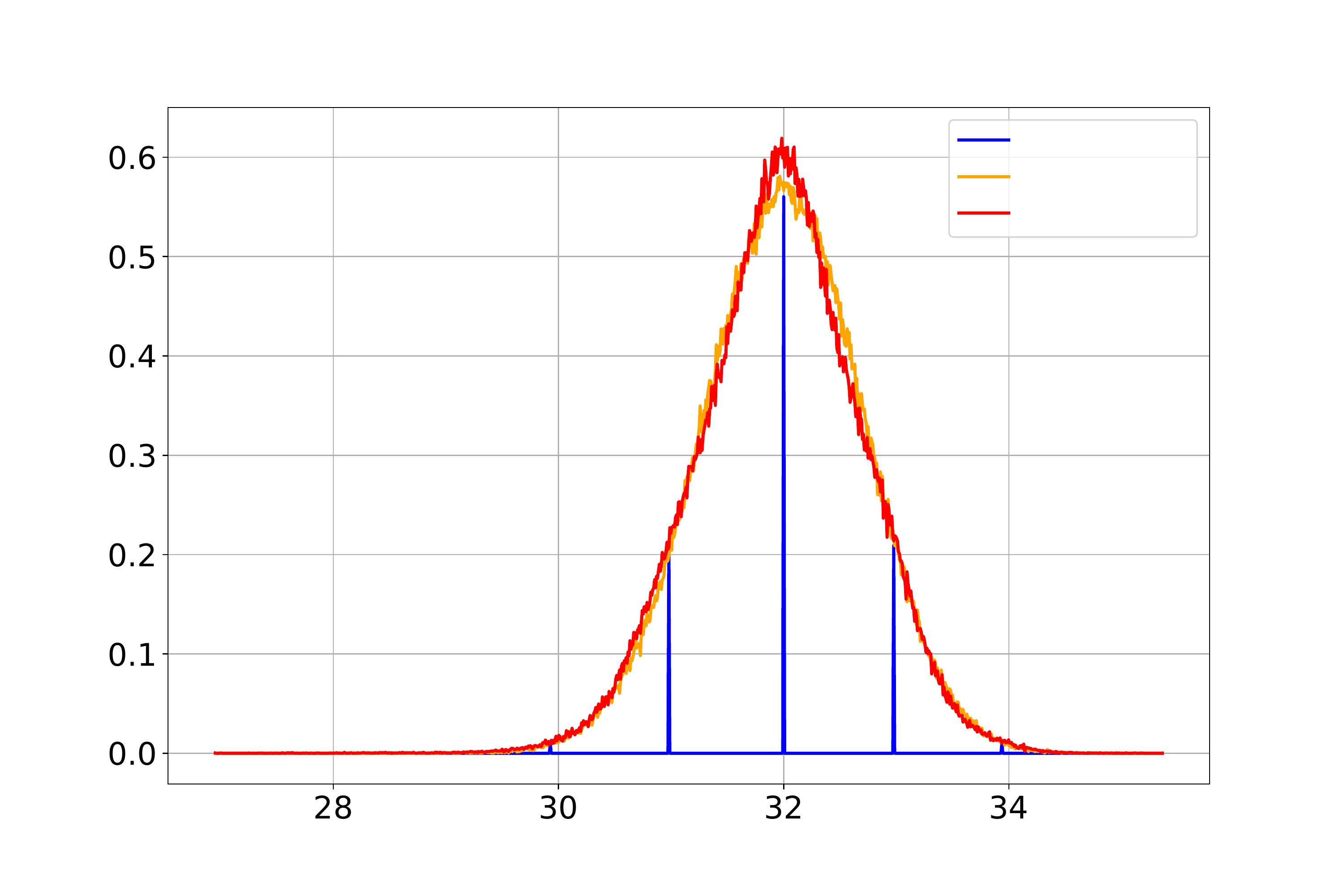}
    \put(-45,116){{\fontsize{5}{6}\selectfont RM(9,2)}}
    \put(-45,111){{\fontsize{5}{6}\selectfont Gaussian}}
    \put(-45,106){{\fontsize{5}{6}\selectfont \KO(9,2)}}
    \put(-180,0){\footnotesize Pairwise distance between two codewords}
    \put(-203,63){\rotatebox[origin=t]{90}{\footnotesize Normalized count}}
    \caption{Histogram of pairwise distances between codewords of the $\KO(9,2)$ code shows a strong resemblance to that of the Gaussian codebook, unlike the classical Reed-Muller code $\RM(9,2)$.}
    \label{fig:2}
    \vspace{-0.08in}
\end{figure}
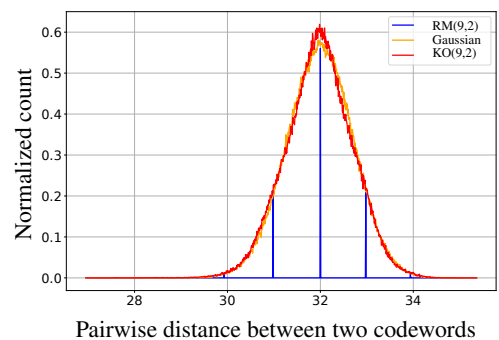

In summary, we make the following contributions:
We introduce novel neural network architectures for the (encoder, decoder) pair that generalizes the Kronecker operation central to RM/Polar codes. We propose training methods that discover novel non-linear codes when  trained over AWGN and provide empirical results showing that this family of non-linear codes improves significantly upon the baseline code it was built on (both RM and Polar codes) whilst having the same encoding and decoding complexity. Interpreting the pairwise distances of the discovered codewords reveals that a \KO~code mimics the distribution of codewords from the random Gaussian codebook, which is known to be reliable but computationally challenging to decode. 
The decoding complexities of \KO~codes are
$O(n\log n)$ where $n$ is the  block length, matching that of efficient decoders for \RM~and Polar codes.  


We highlight that the design principle of KO codes  serves as a general recipe to discover new family of non-linear codes improving upon their linear counterparts. In particular, the construction is not restricted to a specific decoding algorithm, such as successive cancellation (SC). In this paper, we focus on the SC decoding algorithm since it is one of  the most efficient decoders for the RM and Polar family. At this decoding complexity, i.e. $O(n \log n)$, our results demonstrate that we achieve significant gain over these codes. Our preliminary results show that KO codes achieve similar gains over the RM codes, when both are decoded with list-decoding.  We refer to \S\ref{sec:gains_list} for more details.
Designing KO-inspired codes to improve upon the  RPA decoder for RM codes (with complexity $O(n^r \log n)$ \cite{ye2020recursive}), and the  list-decoded Polar codes (with complexity $O(L n \log n)$ \cite{tal2015list}) where $L$ is the list size, are promising active research  directions, and outside the scope of this paper.




\section{Problem formulation  and background}
\label{sec:prob_formul}

We formally define the channel coding problem and provide 
background on Reed-Muller codes, the inspiration for our approach. Our  notation is the following. 
We denote Euclidean vectors by bold face letters like $\bma, \bL$, etc. For $\bL \in \reals^n$, $\bL_{k:m} \define (L_k,\ldots, L_m)$. If $\bv \in \{0,1\}^n$, we define the operator $\oplus_{\bv}$ as $\bx \oplus_{\bv} \by \define \bx + (-1)^{\bv} \by $.


\subsection{Channel coding} 
Let $\bma = (m_1,\ldots, m_k) \in \{0,1\}^k$ denote a block of {\em information/message bits} that we want to transmit. An encoder $g_{\theta}(\cdot)$ is a function parametrized by $\theta$ that maps these information bits into a binary vector $\boldsymbol{x}$ of length $n$, \ie $\boldsymbol{x}= g_{\theta}(\bma) \in \{0, 1\}^n$. The {\em rate} $\rho=k/n$ of such a code measures how many bits of information we are sending per channel use. These codewords are transformed into real (or complex) valued signals, called modulation, before being transmitted  over a channel. For example, Binary Phase Shift Keying (BPSK) modulation maps each $x_i\in\{0,1\}$ to $1-2x_i \in \{\pm 1\}$ up to a universal scaling constant for all $i\in[n]$. Here, we do not strictly separate encoding from modulation and refer to both binary encoded symbols and real-valued transmitted symbols as {\em codewords}. The codewords also satisfy either a hard or soft power constraint. Here we consider the hard power constraint, i.e., $\norm{x}^2=n$.

Upon transmission of this codeword $\boldsymbol{x}$ across a noisy channel $P_{Y|X}(\cdot|\cdot)$, we receive its corrupted version $\boldsymbol{y} \in \reals^n$. The decoder $f_{\phi}(\cdot)$ is a function parametrized by $\phi$ that subsequently processes the received vector $\boldsymbol{y}$ to estimate the information bits $\hat{\bma}=f_{\phi}(\boldsymbol{y})$. The closer $\hat{\bma}$ is to $\bma$, the more reliable the transmission. An error metric, such as Bit-Error-Rate (BER) or Block-Error-Rate (BLER), gauges the performance of the encoder-decoder pair $(g_\theta, f_\phi)$. Note that BER is defined as $\mathrm{BER} \triangleq (1/k)\sum_i \prob{\hat{m}_i \neq m_i}$, whereas $\mathrm{BLER} \triangleq \prob{\hat{\bma} \neq \bma}$.

The design of good codes given a channel  and a fixed set of code parameters $(k,n)$ can be  formulated as: 
\begin{eqnarray}
(\theta,\phi) \;\;\in\;\; \arg \min_{\theta, \phi}\;  \mathrm{BER}(g_\theta, f_\phi)\;,
\label{eq:loss}
\end{eqnarray} 
which is a joint classification problem for $k$ binary classes, and we train on the surrogate loss of cross entropy to make the objective differentiable. While classical optimal codes such as Turbo, LDPC, and Polar codes all have {\em linear} encoders, appropriately parametrizing both the encoder $g_\theta(\cdot)$ and the decoder $f_\phi(\cdot)$ by neural networks (NN) allows for a much broader class of codes, especially non-linear codes. However, 
in the absence of any structure, NNs fail to learn non-trivial codes and end up performing worse than simply repeating each message bit $n/k$ times  \cite{kim2018communication, jiang2019turbo}. 

A fundamental question in machine learning for channel coding is thus: how do we design architectures for our neural encoders and decoders that give the appropriate inductive bias? To gain intuition towards addressing this, we focus on  Reed-Muller (RM) codes. In \S\ref{sec:neural_ko}, we present a novel family of non-linear codes, {\em \KO~codes}, that strictly generalize and improve upon RM codes by capitalizing on their inherent recursive structure. Our approach seamlessly generalizes to Polar codes, explained in \S\ref{sec:polar}. 

\subsection{Reed-Muller (RM) codes}
\label{sec:plotkin_RM}

We use a small example of $\RM(3,1)$ and refer to Appendix~\ref{sec:rm_82} for the larger example in our main results.  

{\bf Encoding.} RM codes are a family of codes parametrized by a variable size $m\in{\mathbb Z}_+$ and an order $r\in{\mathbb Z}_+$ with $r \leq m$,  denoted as $\RM(m,r)$.
It is  defined by an {\em encoder}, which maps binary information bits $\bma \in\{0,1\}^k$ to codewords 
$\boldsymbol{x}\in \{0,1\}^n$. $\RM(m,r)$ code sends $k=\sum_{i=0}^r {m \choose i}$ information bits with $n=2^m$ transmissions. The {\em code distance} measures the minimum distance between all (pairs of) codewords. \prettyref{tab:RM} summarizes these parameters.


\begin{table}[h]
{\tabulinesep=1mm
   \begin{tabu} { c|c|c|c }
   Code length & Code dimension & Rate & Distance \\\hline\hline
 $n=2^m$ & $k=\sum_{i=0}^r {m \choose i}$ & $\rho\!=\!k/n$ & $d\!=\!2^{m-r}$
\end{tabu}}
\vspace{-0.15cm}
\caption{Parameters of a $\RM(m,r)$ code }
\label{tab:RM}
\end{table}

One way to define $\RM(m,r)$ code is via the recursive application of a {\em Plotkin construction}. The basic building block is a mapping $\mathrm{Plotkin}:\{0,1\}^\ell\times\{0,1\}^\ell\to\{0,1\}^{2\ell}$, where
\begin{eqnarray}
    \mathrm{Plotkin}(\boldsymbol{u},\boldsymbol{v})= (\boldsymbol{u} , \boldsymbol{u}\oplus \boldsymbol{v} )\;, \label{eq:def_plotkin}
\end{eqnarray}
with $\oplus$ representing a coordinate-wise XOR and $(\cdot , \cdot)$ denoting concatenation of two vectors \cite{plotkin60}. 
 

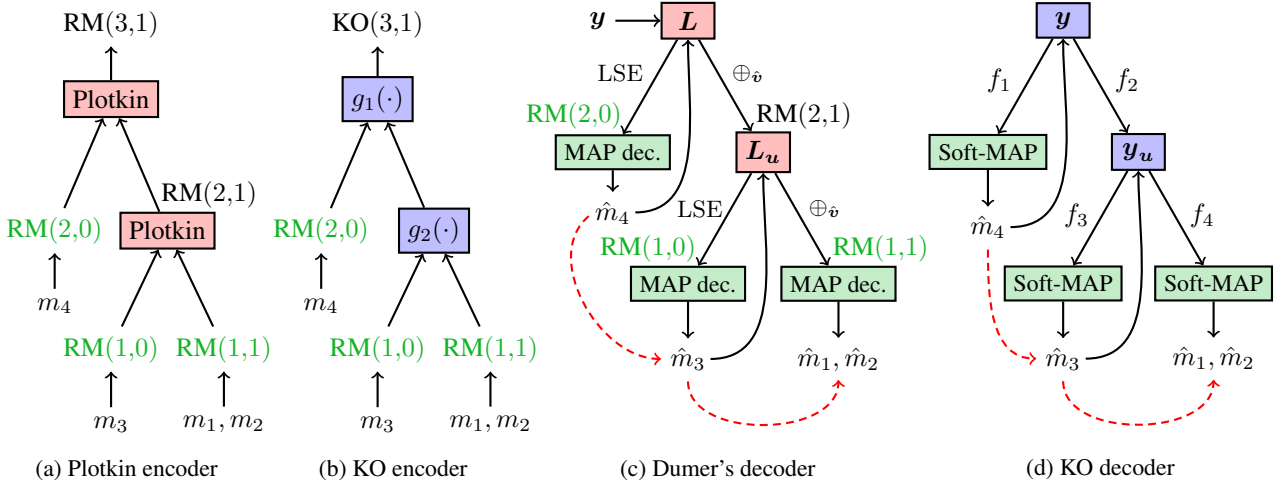
\begin{figure*}[t]
\begin{subfigure}{0.20\textwidth}
	\centering
	\begin{tikzpicture}
	\node at (1,10) (c) {$\RM(3,\!1)$};
	\node [block, fill=red!25] at (1,9) (r1) {Plotkin};
	\node at (0.25,7.25) (r2) {\textcolor{MyGreen1}{$\RM(2,\!0)$}};
	\node [block, fill=red!25] at (1.75,7.25) (r3) {Plotkin};
	\node at (1,5.675) (r4) {\textcolor{MyGreen1}{$\RM(1,\!0)$}};
	\node at (2.5,5.675) (r5) {\textcolor{MyGreen1}{$\RM(1,\!1)$}};
	
	\node at (0.25,6.25) (m4) {{$m_4$}};
	\node at (1,4.675) (m3) {{$m_3$}};
	\node at (2.5,4.675) (m12) {{$m_1,m_2$}};
	
	\draw [->,thick] (r1)--(c);
	
	\draw [->,thick] (r2)--(r1);
	\draw [->,thick] (r3)--(r1);
	\draw [->,thick] (r4)--(r3);
	\draw [->,thick] (r5)--(r3);
	
	\draw [->,thick] (m4)--(r2);
	\draw [->,thick] (m3)--(r4);
	\draw [->,thick] (m12)--(r5);
	\end{tikzpicture}
	\put(-45,91){$\RM(2,\!1)$}
	\caption{Plotkin encoder
	}
	\label{Fig_RMEnc31}
\end{subfigure}
\hfill
\begin{subfigure}{0.20\textwidth}
	\centering
	\begin{tikzpicture}
	\node at (1,10) (c) {$\KO(3,\!1)$};
	\node [block, fill=blue!25] at (1,9) (r1) {$g_1(\cdot)$};
	\node at (0.25,7.25) (r2) {\textcolor{MyGreen1}{$\RM(2,\!0)$}};
	\node [block, fill=blue!25] at (1.75,7.25) (r3) {$g_2(\cdot)$};
	\node at (1,5.675) (r4) {\textcolor{MyGreen1}{$\RM(1,\!0)$}};
	\node at (2.5,5.675) (r5) {\textcolor{MyGreen1}{$\RM(1,\!1)$}};
	
	\node at (0.25,6.25) (m4) {{$m_4$}};
	\node at (1,4.675) (m3) {{$m_3$}};
	\node at (2.5,4.675) (m12) {{$m_1,m_2$}};
	
	\draw [->,thick] (r1)--(c);
	
	\draw [->,thick] (r2)--(r1);
	\draw [->,thick] (r3)--(r1);
	\draw [->,thick] (r4)--(r3);
	\draw [->,thick] (r5)--(r3);
	
	\draw [->,thick] (m4)--(r2);
	\draw [->,thick] (m3)--(r4);
	\draw [->,thick] (m12)--(r5);
	\end{tikzpicture}
	\caption{KO encoder
	}
	\label{Fig_NEnc31}
\end{subfigure}
\hfill
\begin{subfigure}{0.29\textwidth}
	\centering
	\begin{tikzpicture}
	\node at (-0.2,9) (y) {$\boldsymbol{y}$};
	\node [block, fill=red!25] at (1,9) (r1) {$\boldsymbol{L}$};
	\node [block, fill=MyGreen1!25] at (0,7.25) (r2) {\small{MAP dec.}};
	\node [block, fill=red!25] at (2,7.25) (r3) {$\boldsymbol{L}_{\boldsymbol{u}}$};
	\node [block, fill=MyGreen1!25] at (1,5.5) (r4) {\small{MAP dec.}};
	\node [block, fill=MyGreen1!25] at (3,5.5) (r5) {\small{MAP dec.}};
	
	\node at (0,6.45) (m4) {{$\hat{m}_4$}};
	\node at (1,4.5) (m3) {{$\hat{m}_3$}};
	\node at (3,4.5) (m12) {{$\hat{m}_1,\hat{m}_2$}};
	
	\node at (0.1,8.3) (LSE) {\small{$\mathrm{LSE}$}};
	\node at (1.8,8.3) (Plus) {\small{$\oplus_{\hat{\bv}}$}};

	\node at (1.16, 6.5) (f3) {\small{$\mathrm{LSE}$}};
	\node at (2.82, 6.5) (f4) {\small{$\oplus_{\hat{\bv}}$}};

	\draw [->,thick] (y)--(r1);
	
	\draw [->,thick] (r1)--(r2);
	\draw [->,thick] (r1)--(r3);
	\draw [->,thick] (r3)--(r4);
	\draw [->,thick] (r3)--(r5);
	
	\draw [->,thick] (r2)--(m4);
	\draw [->,thick] (r4)--(m3);
	\draw [->,thick] (r5)--(m12);
	
	\draw [->,densely dashed,thick,red] (-0.32,6.4) to [out=220,in=180] (m3.west);
	\draw [->,densely dashed,thick,red] (1,4.2) to [out=-90,in=-90] (m12.south);
	
	\draw [->,thick] (0.3,6.45) to [out=0,in=-90] (r1.south);
	\draw [->,thick] (1.3,4.5) to [out=0,in=-90] (r3.south);

	\end{tikzpicture}
    \put(-140,121){\textcolor{MyGreen1}{ $\RM(2,\!0)$ }}
    \put(-53,121){\textcolor{black}{ $\RM(2,\!1)$ }} 
    \put(-112,70.5){\textcolor{MyGreen1}{ $\RM(1,\!0)$ }}
    \put(-24,70.5){\textcolor{MyGreen1}{ $\RM(1,\!1)$ }}
	\caption{Dumer's decoder}
	\label{Fig_Dumer_decoder31}
\end{subfigure}
\hfill
\begin{subfigure}{0.29\textwidth}
	\centering
	\begin{tikzpicture}
	\node [block, fill=blue!25] at (1,9) (r1) {$\boldsymbol{y}$};
	\node [block, fill=MyGreen1!25] at (0,7.25) (r2) {\small{Soft-MAP}};
	\node [block, fill=blue!25] at (2,7.25) (r3) {$\boldsymbol{y}_{\boldsymbol{u}}$};
	\node [block, fill=MyGreen1!25] at (1,5.5) (r4) {\small{Soft-MAP}};
	\node [block, fill=MyGreen1!25] at (3,5.5) (r5) {\small{Soft-MAP}};
	
	\node at (0,6.25) (m4) {{$\hat{m}_4$}};
	\node at (1,4.5) (m3) {{$\hat{m}_3$}};
	\node at (3,4.5) (m12) {{$\hat{m}_1,\hat{m}_2$}};
	
	\node at (0.15,8.2) (f1) {$f_1$};
	\node at (1.85,8.2) (f2) {$f_2$};
	
	\node at (1.18,6.4) (f3) {$f_3$};
	\node at (2.85,6.4) (f4) {$f_4$};
	
	
	\draw [->,thick] (r1)--(r2);
	\draw [->,thick] (r1)--(r3);
	\draw [->,thick] (r3)--(r4);
	\draw [->,thick] (r3)--(r5);
	
	\draw [->,thick] (r2)--(m4);
	\draw [->,thick] (r4)--(m3);
	\draw [->,thick] (r5)--(m12);
	
	\draw [->,densely dashed,thick,red] (0,5.95) to [out=-90,in=180] (m3.west);
	\draw [->,densely dashed,thick,red] (1,4.2) to [out=-90,in=-90] (m12.south);
	
	\draw [->,thick] (0.3,6.25) to [out=0,in=-90] (r1.south);
	\draw [->,thick] (1.3,4.5) to [out=0,in=-90] (r3.south);

	\end{tikzpicture}
     \caption{KO decoder
	}
	\label{Fig_NN_decoder31}
\end{subfigure}
\caption{ Plotkin trees for $\RM(3,1)$ and $\KO(3,1)$ codes; Leaves are shown in green. Red arrows indicate the bit decoding order.}
\end{figure*}

In view of the Plotkin construction, RM codes are recursively defined as a set of codewords of the form: 
\begin{align}
    \RM(m,r) &= \{ (\boldsymbol{u}, \boldsymbol{u}\oplus \boldsymbol{v}) :\boldsymbol{u} \in \RM(m-1,r), \nonumber \\
    & \hspace{4em} \boldsymbol{v} \in \RM(m-1,r-1)\},
    \label{eq:RM}
\end{align}
where $\RM(m,0)$ is a repetition code that repeats a single information bit $2^m$ times, i.e., $\boldsymbol{x}=(m_1,m_1,\ldots,m_1)$. When $r=m$, the full-rate $\RM(m,m)$ code is also recursively defined as a Plotkin construction of two $\RM(m-1,m-1)$ codes. Unrolling the recursion in Eq.~\eqref{eq:RM}, a $\RM(m,r)$ encoder can be 
represented by a corresponding (rooted and binary) computation tree, which we refer to as its {\em Plotkin tree}. 
In this tree, each branch represents a Plotkin mapping of two codes of appropriate lengths, recursively applied from the leaves to the root.

Figure~\ref{Fig_RMEnc31} 
illustrates such a Plotkin tree decomposition of $\RM(3,1)$ encoder. Encoding starts from the bottom right leaves. The leaf $\RM(1,0)$ maps $m_3$ to $(m_3,m_3)$ (repetition), and 
another leaf $\RM(1,1)$ maps $(m_1,m_2)$ to $(m_1,m_1 \oplus m_2)$ (Plotkin mapping of two $\RM(0,0)$ codes). Each branch in this tree performs the Plotkin construction of Eq.~\eqref{eq:def_plotkin}. The next operation is the  parent of these two leaves, which performs ${\rm Plotkin}(\RM(1,1),\RM(1,0)) = {\rm Plotkin}( (m_1,m_1 \oplus m_2),(m_3,m_3) )$ which outputs the vector $(m_1, m_1 \oplus m_2,m_1\oplus m_3,  m_1 \oplus m_2 \oplus m_3)$, which is known as $\RM(2,1)$ code. This coordinate-wise Plotkin construction is applied recursively one more time to combine $\RM(2,0)$ and $\RM(2,1)$ at the root of the tree. The resulting codewords are $\RM(3,1)={\rm Plotkin}(\RM(2,1),\RM(2,0)
) = {\rm Plotkin}( (m_1,m_1 \oplus m_2,m_1\oplus m_3, m_1 \oplus m_2\oplus m_3), (m_4,m_4,m_4,m_4))$.

This recursive structure of RM codes $(i)$ inherits the good minimum distance property of the Plotkin construction and $(ii)$ enables
efficient decoding. 

{\bf Decoding.} Since 
\cite{reed1954class}, 
there have been several decoders for RM codes; \cite{abbe2020reedmuller} is a detailed survey. 
We focus on the most efficient one, called {\em Dumer's recursive decoding} \citep{dumer2004recursive, dumer2006soft1, dumer2006soft2} that fully capitalizes on the recursive Plotkin construction in Eq.~\eqref{eq:RM}. The basic principle is: to decode an RM codeword $\boldsymbol{x}=(\boldsymbol{u},\boldsymbol{u} \oplus \boldsymbol{v}) \in \RM(m,r)$, we first recursively decode the left sub-codeword $\boldsymbol{v} \in \RM(m-1, r-1)$ and then the right sub-codeword $\boldsymbol{u} \in \RM(m-1, r)$, and we use them together to stitch back the original codeword. This recursion is continued until we reach the leaf nodes, where we perform maximum a posteriori (MAP) decoding. Dumer's recursive decoding is also referred to as \textit{successive cancellation} decoding in the context of polar codes \cite{arikan2009channel}.

Figure~\ref{Fig_Dumer_decoder31} illustrates this decoding procedure for $\RM(3,1)$. Dumer's decoding starts at the root and uses the soft-information of codewords to decode the message bits. Suppose that the message bits $\bma = (m_1,\ldots, m_4)$ are encoded into an $\RM(3,1)$ codeword $\bx \in \{0, 1\}^8$ using the Plotkin encoder in Figure~\ref{Fig_RMEnc31}. Let $\by \in \reals^8$ be the corresponding noisy codeword received at the decoder.
To decode the bits $\bma$, we first obtain the soft-information of the codeword $\bx$, \ie, we compute its Log-Likehood-Ratio (LLR) $\boldsymbol{L} \in \reals^8$:
\begin{align*}
L_i = \log \frac{\prob{ y_i |x_i=0}}{\prob{ y_i |x_i=1}}, \quad i=1,\ldots,8.
\end{align*}
We next use $\bL$ to compute soft-information for its left and right children: the $\RM(2,0)$ codeword $\bv$ and the $\RM(2,1)$ codeword $\bu$. We start with the left child $\bv$.

Since the codeword  $\bx = (\bu, \bu \oplus \bv)$, we can also represent its left child as $\bv = \bu \oplus (\bu \oplus \bv) = \bx_{1:4} \oplus \bx_{5:8}$. Hence its LLR vector $\bL_{\bv} \in \reals^4$ can be readily obtained from that of $\bx$. In particular it is given by the log-sum-exponential transformation: $\bL_{\bv} = \mathrm{LSE}(\bL_{1:4}, \bL_{5:8})$, where $\mathrm{LSE}(a,b) \define \log((1+e^{a+b})/(e^a + e^b))$ for $a,b \in \reals$. Since this feature $\bL_{\bv}$ corresponds to a repetition code, $\bv = (m_4, m_4, m_4, m_4)$, majority decoding (same as the MAP) on the sign of $\bL_{\bv}$ yields the decoded message bit as $\hat{m}_4$. Finally, the left codeword is decoded as  $\hat{\bv}=(\hat{m}_4, \hat{m}_4, \hat{m}_4, \hat{m}_4)$.

Having decoded the left $\RM(2,0)$ codeword $\hat{\bv}$, our goal is to now obtain soft-information $\bL_{\bu} \in \reals^4$ for the right $\RM(2,1)$ codeword $\bu$. Fixing $\bv = \hat{\bv}$, notice that the codeword $\bx= (\bu, \bu \oplus \hat{\bv})$ can be viewed as a $2$-repetition of $\bu$ depending on the parity of $\hat{\bv}$. Thus the LLR $\bL_{\bu}$ is given by LLR addition accounting for the parity of $\hat{\bv}$: $\bL_{\bu} =  \bL_{1:4} \oplus_{\hat{\bv}} \bL_{5:8} = \bL_{1:4} + (-1)^{\hat{\bv}} \bL_{5:8}$. Since $\RM(2,1)$ is an internal node in the tree, we again recursively decode its left child $\RM(1,0)$ and its right child $\RM(1,1)$, which are both leaves. For $\RM(1,0)$, decoding is similar to that of $\RM(2,0)$ above, and we obtain its information bit $\hat{m_3}$ by first applying the log-sum-exponential function on the feature $\bL_{\bu}$ and then majority decoding. Likewise, we obtain the LLR feature $\bL_{\bu \bu} \in \reals^2$ for the right $\RM(1,1)$ child using parity-adjusted LLR addition on $\bL_{\bu}$. Finally, we decode its corresponding bits $(\hat{m}_1, \hat{m}_2)$ using efficient MAP-decoding of first order RM codes \cite{abbe2020reedmuller}. Thus we obtain the full block of decoded message bits as $\hat{\bma} = (\hat{m}_1, \hat{m}_2,\hat{m}_3, \hat{m}_4)$. 

An important observation from Dumer's algorithm is that the sequence of bit decoding in the tree is: $\RM(2,0) \rightarrow \RM(1,0) \rightarrow \RM(1,1)$. A similar decoding order holds for all $\RM(m, 2)$ codes, where all the left leaves (order-$1$ codes) are decoded first from top to bottom, and the right-most leaf (full-rate $\RM(2,2)$) is decoded at the end.





\section{\KO~codes: Novel Neural codes} 
\label{sec:neural_ko}
We design \KO~codes  
using the Plotkin tree as the skeleton of a new neural network architecture, which 
strictly improve upon their classical counterparts.

%

\noindent {\bf \KO~encoder}. Earlier we saw the design of RM codes via recursive Plotkin mapping. Inspired by this elegant construction, we present a new family of codes, called \emph{ \KO~codes}, denoted as \KO$(m, r, g_\theta, f_\phi)$. These codes are parametrized by a set of four parameters: a non-negative integer pair $(m,r)$, a finite set of encoder neural networks $g_\theta$, and a finite set of decoder neural networks $f_\phi$. In particular, for any fixed pair $(m,r)$, our KO encoder inherits the same code parameters $(k,n,\rho)$ and the same Plotkin tree skeleton of the RM encoder. However, a critical distinguishing component of our $\KO(m,r)$ encoder is a set of encoding neural networks $g_\theta = \{g_i\}$ that strictly generalize the Plotkin mapping: to each internal node $i$ of the Plotkin tree, we associate a neural network $g_i$ that applies a coordinate-wise real valued non-linear mapping $(\boldsymbol{u},\boldsymbol{v}) \mapsto g_i(\boldsymbol{u},\boldsymbol{v}) \in \reals^{2\ell}$ as opposed to the classical binary valued Plotkin mapping $(\boldsymbol{u},\boldsymbol{v}) \mapsto (\bu, \boldsymbol{u} \oplus \boldsymbol{v}) \in \{0, 1 \}^{2\ell}$. Figure~\ref{Fig_NEnc31} illustrates this for the $\KO(3,1)$ encoder.

The significance of our KO encoder $g_\theta$ is that by allowing for general nonlinearities $g_i$ to be learnt at each node we enable for a much richer and broader class of nonlinear encoders and codes to be discovered on a whole, which contribute to non-trivial gains over standard RM codes. Further, we have the same encoding complexity as that of an RM encoder since each $g_i: {\mathbb R}^{2} \to {\mathbb R}$ is applied coordinate-wise on its vector inputs. The parameters of these neural networks $g_i$ are trained via stochastic gradient descent on the cross entropy loss. See \S\ref{sec:exp_detail} for experimental detailas. 



\noindent {\bf \KO~decoder}. Training the encoder is possible only if we have a corresponding decoder. This necessitates the need for an efficient family of matching decoders. Inspired by the  Dumer's  decoder, we present a new family of \emph{KO decoders} that fully capitalize on the recursive structure of KO encoders via the Plotkin tree. 

Our KO decoder has three distinct features: 
{$(i)$ Neural decoder}: The KO decoder architecture is parametrized by a set of decoding neural networks $f_\phi=\{ (f_{2i-1}, f_{2i}) \}$. Specifically, to each internal node $i$ in the tree, we associate $f_{2i-1}$ to its left branch whereas $f_{2i}$ corresponds to the right branch. Figure~\ref{Fig_NN_decoder31} shows this for the $\KO(3,1)$ decoder. The pair of decoding neural networks $(f_{2i-1}, f_{2i})$ can be viewed as matching decoders for the corresponding encoding network $g_i$: While $g_i$ encodes the left and right codewords arriving at this node, the outputs of $f_{2i-1}$ and $f_{2i}$ represent appropriate Euclidean feature vectors for decoding them. Further, $f_{2i-1}$ and $f_{2i}$ can also be viewed as a generalization of Dumer's decoding to nonlinear real codewords: $f_{2i-1}$ generalizes the $\mathrm{LSE}$ function, while $f_{2i}$ extends the operation $\oplus_{\hat{\bv}}$. Note that both the functions $f_{2i-1}$ and $ f_{2i}$ are also applied coordinate-wise and hence we inherit the same decoding complexity as Dumer's.
{$(ii)$ Soft-MAP decoding}: Since the classical MAP decoding to decode the bits at the leaves is not differentiable, we design a new differentiable counterpart, the \emph{Soft-MAP decoder}. Soft-MAP decoder enables gradients to pass through it, which is crucial for training the neural (encoder, decoder) pair $(g_\theta, f_\phi)$ in an end-to-end manner. {$(iii)$ Channel agnostic}: Our decoder directly operates on the received noisy codeword $\by \in \reals^n$ while Dumer's decoder uses its LLR transformation $\bL \in \reals^n$. Thus, our decoder can learn the appropriate channel statistics for decoding directly from $\by$ alone; in contrast, Dumer's algorithm requires precise channel characterization, which is not usually known. 



\section{Main results} 
\label{sec:experimental_results} 

We train the \KO~encoder $g_\theta$ and \KO~decoder $f_\phi$ from \S\ref{sec:neural_ko} using an approximation of the BER loss in \eqref{eq:loss}. 
The details are provided in \S\ref{sec:exp_detail}. In this section we focus on the second-order $\KO(8,2)$ and $\KO(9,2)$ codes.

\subsection{\KO~codes improve over RM codes} 
\label{sec:awgn}

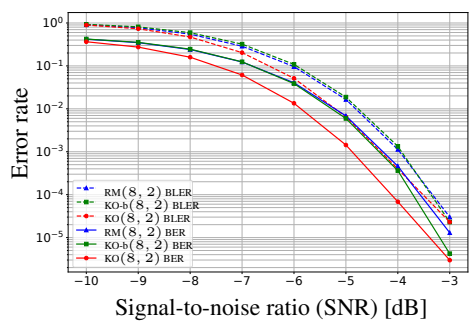
\begin{figure}[H] 
    \centering
    \includegraphics[width=.4\textwidth]{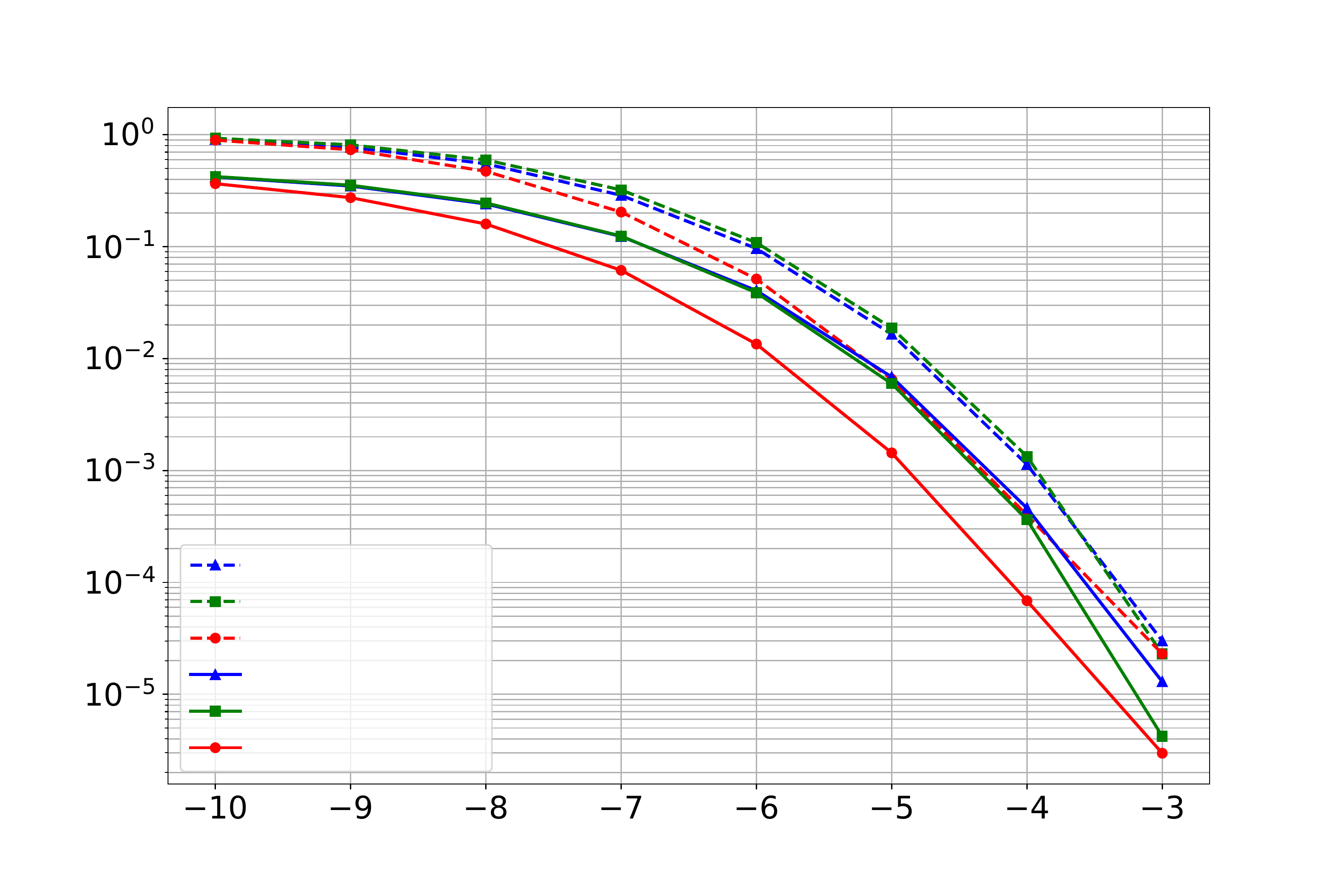}
    \put(-157,45){{\fontsize{4}{5}\selectfont $\RM(8,2)$ BLER}}
    \put(-157,40){\fontsize{4}{5}\selectfont \KO-b$(8,2)$ BLER}
    \put(-157,35){\fontsize{4}{5}\selectfont \KO$(8,2)$ BLER}
    \put(-157,30){\fontsize{4}{5}\selectfont RM$(8,2)$ BER}
    \put(-157,25){\fontsize{4}{5}\selectfont \KO-b$(8,2)$ BER}
    \put(-157,20){\fontsize{4}{5}\selectfont \KO$(8,2)$ BER}
    \put(-153,0){\footnotesize Signal-to-noise ratio (SNR) [dB]}
    \put(-192,63){\rotatebox[origin=t]{90}{\footnotesize Error rate}}
    \vspace{-0.1cm}
\caption{Neural network based $\KO(8,2)$ and \KO-b$(8,2)$ improve upon $\RM(8,2)$ in BER and BLER, but the gain is small for the binarized codewords of \KO-b$(8,2)$ (for all the codes, the code dimension is $37$ and block length is $256$).}
    \label{fig:awgn82}
\end{figure}

In Figure~\ref{fig:1}, the trained $\KO(9,2)$ improves over the competing $\RM(9,2)$ both in BER and BLER. The superiority in BLER is unexpected as  
our training loss is a surrogate for the BER. Though one would prefer to train on BLER as it is more relevant in practice, it is challenging to design a surrogate loss for BLER that is also differentiable:
all literature on learning decoders minimize only BER \cite{kim2020physical,nachmani2018deep,dorner2017deep}.  
Consequently, improvements in BLER with trained encoders and/or decoders are rare. 
We discover a code that improves both BER and BLER, and 
we observe a similar gain with $\KO(8,2)$ in Figure~\ref{fig:awgn82}. Performance of a binarized version \KO-b$(8,2)$ is also shown, which we describe further in \S\ref{sec:ablation_studies}.

\vspace{- 0.8 em}



\subsection{Interpreting \KO~codes}
\label{sec:interp}

We interpret the learned encoders and decoders to explain the source of the performance gain.

\medskip\noindent

{\bf Interpreting the \KO~encoder.}
To interpret the learned \KO~code, we examine the pairwise distance between codewords. In classical linear coding, pairwise distances are expressed in terms of the weight distribution of the code, which counts how many codewords of each specific Hamming weight $1,2,\dots,n$ exist in the code. The weight distribution of linear codes are used to derive analytical bounds, that can be explicitly computed, on the BER and BLER over AWGN channels  \cite{sason2006performance}.  For nonlinear codes, however, the weight distribution does not capture pairwise distances. Therefore, we explore the distribution of all the pairwise distances of non-linear KO codes that can play the same role as the weight distribution does for linear codes. 

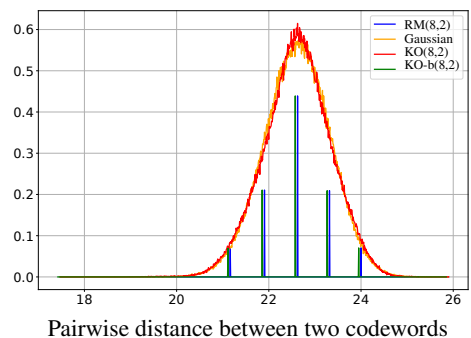
\begin{figure}[H] 
    \centering
    \includegraphics[width=.43\textwidth]{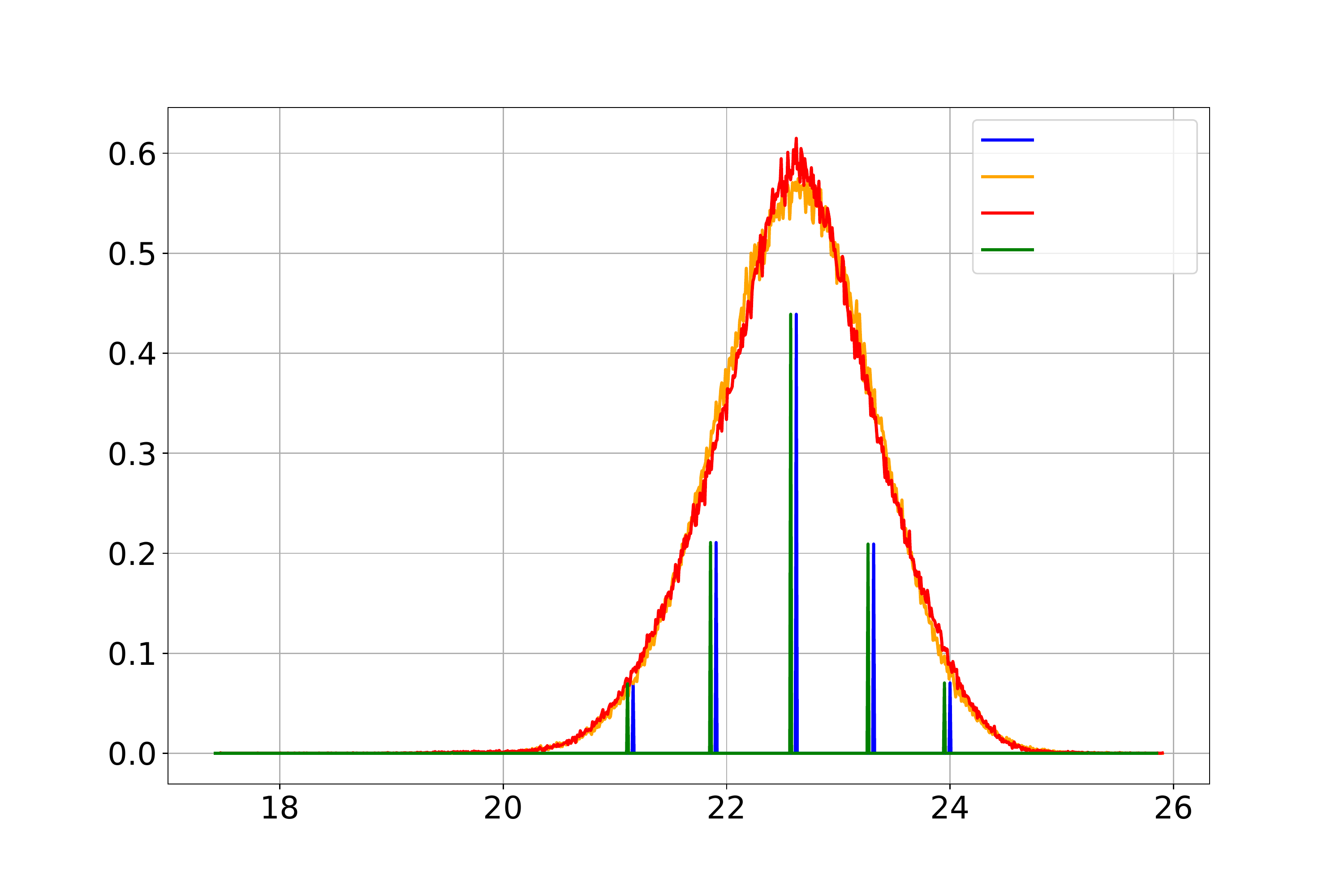}
    \put(-45,116){{\fontsize{5}{6}\selectfont RM(8,2)}}
    \put(-45,111){{\fontsize{5}{6}\selectfont Gaussian}}
    \put(-45,106){{\fontsize{5}{6}\selectfont \KO(8,2)}}
    \put(-45,101){{\fontsize{5}{6}\selectfont \KO-b(8,2)}}
    \put(-180,0){\footnotesize Pairwise distance between two codewords}
    \vspace{-0.1cm}
    \caption{Histograms of pairwise distances between codewords for $(8,2)$ codes reveal that $\KO(8,2)$ code has learned an approximate Gaussian codebook that can be efficiently decoded.}
    \label{fig:enc82}

\end{figure}

The pairwise distance distribution of the RM codes remains an active area of research as it is used to prove that RM codes achieve the capacity \cite{kaufman2012weight,abbe2015reed,sberlo2020performance} (Figure~\ref{fig:enc82} blue). 
However, these results are asymptotic in the block length and do not guarantee a good performance, especially in the small-to-medium block lengths that we are interested in. 
On the other hand, Gaussian codebooks, codebooks randomly picked from the ensemble of all Gaussian codebooks, are known to be asymptotically optimal, i.e., achieving the capacity \cite{shannon1948mathematical}, and also demonstrate optimal finite-length scaling laws closely related to the pairwise distance distribution \cite{polyanskiy2010channel} (Figure~\ref{fig:enc82} orange).

Remarkably, the pairwise distance distribution of \KO~code shows a staggering resemblance to that of the Gaussian codebook of the same rate $\rho$ and blocklength $n$ (Figure~\ref{fig:enc82} red). This is an unexpected phenomenon since we minimize only BER. 
We posit that the NN training has learned to construct a Gaussian-like codebook, in order to minimize BER.
Most importantly, unlike the Gaussian codebook, KO codes constructed via NN training are fully compatible with  efficient decoding. 
This phenomenon is observed for all order-$2$ codes we trained 
(e.g.,~Figure~\ref{fig:2} for $\KO(9,2)$). 

    \vspace{-0.3cm}
\medskip\noindent
{\bf Interpreting the \KO~decoder.} 
We now analyze how the \KO~decoder contributes to the gains in BLER over the RM decoder. 
Let $\bma=(\bma_{(7,1)},\ldots, \bma_{(2,2)})$ denote the block of transmitted message bits, where the ordered set of indices $\calL = \{(7,1),\ldots, (2,2)\}$ correspond to the leaf branches (RM codes) of the Plotkin tree. Let $\hat{\bma}$ be the decoded estimate by the $\KO(8,2)$ decoder.

We provide Plotkin trees of $\RM(8,2)$  and $\KO(8,2)$  decoders in Figures~\ref{Fig_RMDec82} and \ref{Fig_NDec82} in the appendix.
Recall that for this $\KO(8,2)$ decoder, similar to the $\KO(3,1)$ decoder in Figure~\ref{Fig_NN_decoder31}, we decode each sub-code in the leaves sequentially, starting from the $(7,1)$ branch down to $(2,2)$: $\hat{\bma}_{(7,1)}\rightarrow \ldots \rightarrow \hat{\bma}_{(2,2)}$. In view of this decoding order, BLER, defined as $\prob{\hat{\bma} \neq \bma}$, can be decomposed as
\begin{align}
    \!\prob{\hat{\bma} \neq \bma} \!=\! \sum_{i \in \calL} \prob{\hat{\bma}_i \neq \bma_i, \hat{\bma}_{1:i-1} = \bma_{ 1:i-1}}.
    \label{eq:bler_decomp}
\end{align}
In other words, BLER can also be represented as the sum of the fraction of errors the decoder makes in each of the leaf branches when no errors were made in the previous ones. Thus, each term in Eq.~(\ref{eq:bler_decomp}) can be viewed as the contribution of each sub-code to the total BLER. 



\begin{figure}[h]
\centering
\includegraphics[width=0.35\textwidth]{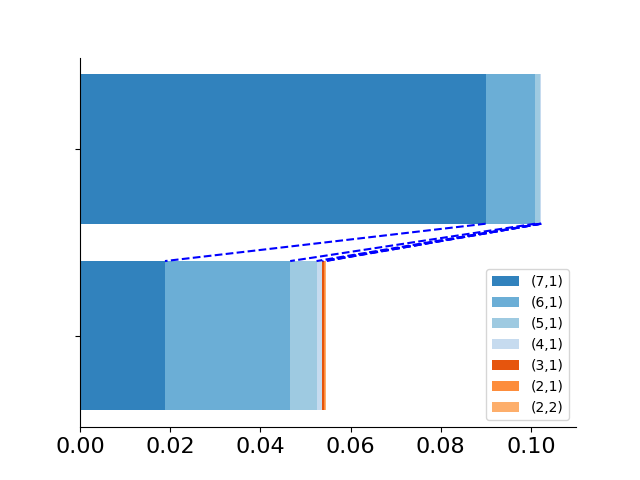}
\put(-188,85){$\RM(8,2)$} 
\put(-188,35){$\KO(8,2)$} 
\put(-100,-5){BLER}
    \caption{Separating each sub-code contribution in the $\KO(8,2)$ decoder and the $\RM(8,2)$ decoder reveals that $\KO(8,2)$ improves in the total BLER by balancing the contributions more evenly over the sub-codes.}
    \label{fig:dec82}
\end{figure}

This is plotted in \prettyref{fig:dec82}, which shows that the $\KO(8,2)$ decoder achieves better BLER than the $\RM(8,2)$ decoder by making major gains in the leftmost $(7,1)$ branch (which is decoded first) at the expense of other branches. However, the decoder (together with the encoder) has learnt to better balance these contributions evenly across all branches, resulting in lower BLER overall. The unequal  errors in the branches of the RM code has been observed before, and some efforts made to balance them \cite{dumer2001near}; that \KO\ codes learn such a balancing scheme purely from data  is, perhaps, remarkable. 

\subsection{Robustness to non-AWGN channels}
\label{sec:robust_adapt}

As the environment changes dynamically in real world channels, robustness is crucial in practice. We therefore test the \KO~code under canonical channel models and demonstrate robustness, i.e.,~the ability of a code trained on AWGN to perform well under a different channel {\em without retraining}. 
It is well known that Gaussian noise is the worst case noise among all noise with the same variance \cite{lapidoth1996nearest,shannon1948mathematical} when an optimal decoder is used, which might take an exponential time. When decoded with efficient decoders, as we do with both RM and \KO~codes, catastrophic failures have been reported in the case of Turbo decoders \cite{kim2018communication}. We show that both RM codes and \KO~codes are robust  and that \KO~codes maintain their gains over RM codes as the channels vary. 

\vspace{-0.4cm}
\begin{figure}[H] 
    \centering
    \includegraphics[width=.4\textwidth]{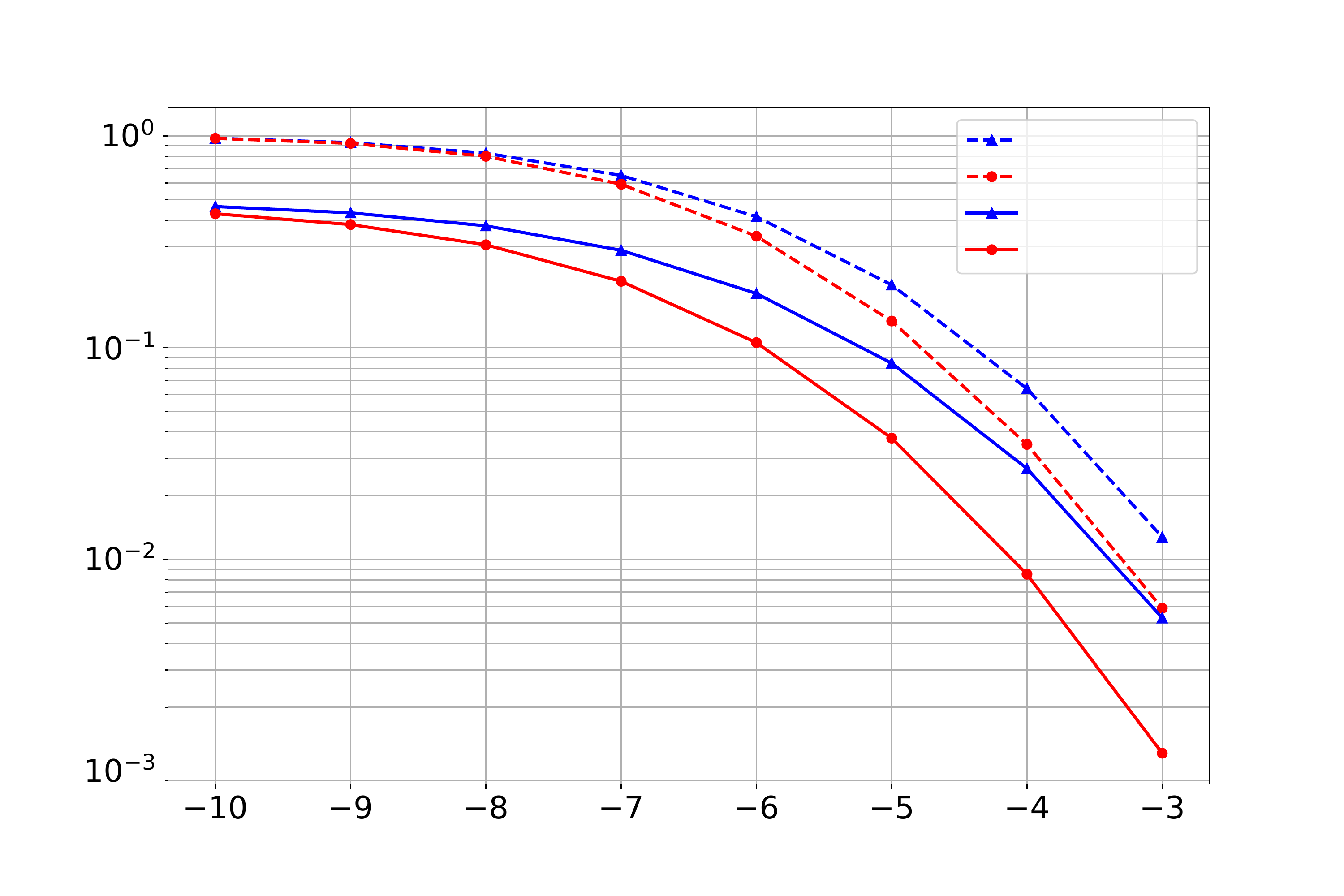}
    \put(-45,108){{\fontsize{5}{6}\selectfont RM BLER}}
    \put(-45,103){{\fontsize{5}{6}\selectfont\KO~BLER}}
    \put(-45,98){{\fontsize{5}{6}\selectfont RM BER}}
    \put(-45,93){{\fontsize{5}{6}\selectfont \KO~BER}}
    \put(-153,0){\footnotesize Signal-to-noise ratio (SNR) [dB]}
    \put(-192,63){\rotatebox[origin=t]{90}{\footnotesize Error rate}}
    \caption{$\KO(8,2)$ trained on AWGN is robust when tested on a fast fading channel and maintains a significant gain over RM(8,2).}
    \label{fig:ff}
  \vspace{-0.3cm}
\end{figure}
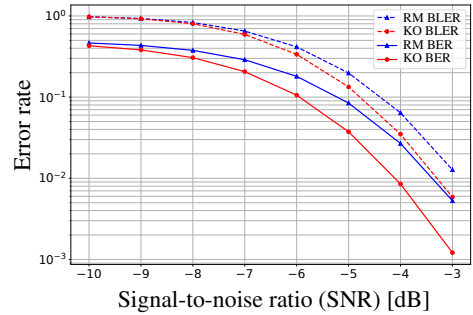

We first test on a {\em Rayleigh fast fading channel}, defined as $y_i=a_i x_i+n_i$, where $x_i$ is the transmitted symbol, $y_i$ is the received symbol, $n_i\sim {\cal N}(0, \sigma^2)$ is the additive Gaussian noise, and $a$ is from a Rayleigh distribution with the variance of $a$ chosen as ${\Expect[a_i^2]=1}$.

We next test on a bursty channel,  defined as $y_i= x_i+n_i+w_i$, where $x_i$ is the input symbol, $y_i$ is the received symbol, $n_i\sim {\cal N}(0, \sigma^2)$ is the additive Gaussian noise, and $w_i\sim {\cal N}(0, \sigma_b^2)$ with probability $\rho$ and $w_i=0$ with probability $1-\rho$. In the experiment, we choose $\rho=0.1$ and $\sigma_b=\sqrt{2}\sigma$.
 
\vspace{-0.4cm}
\begin{figure}[H] 
    \centering
    \includegraphics[width=.4\textwidth]{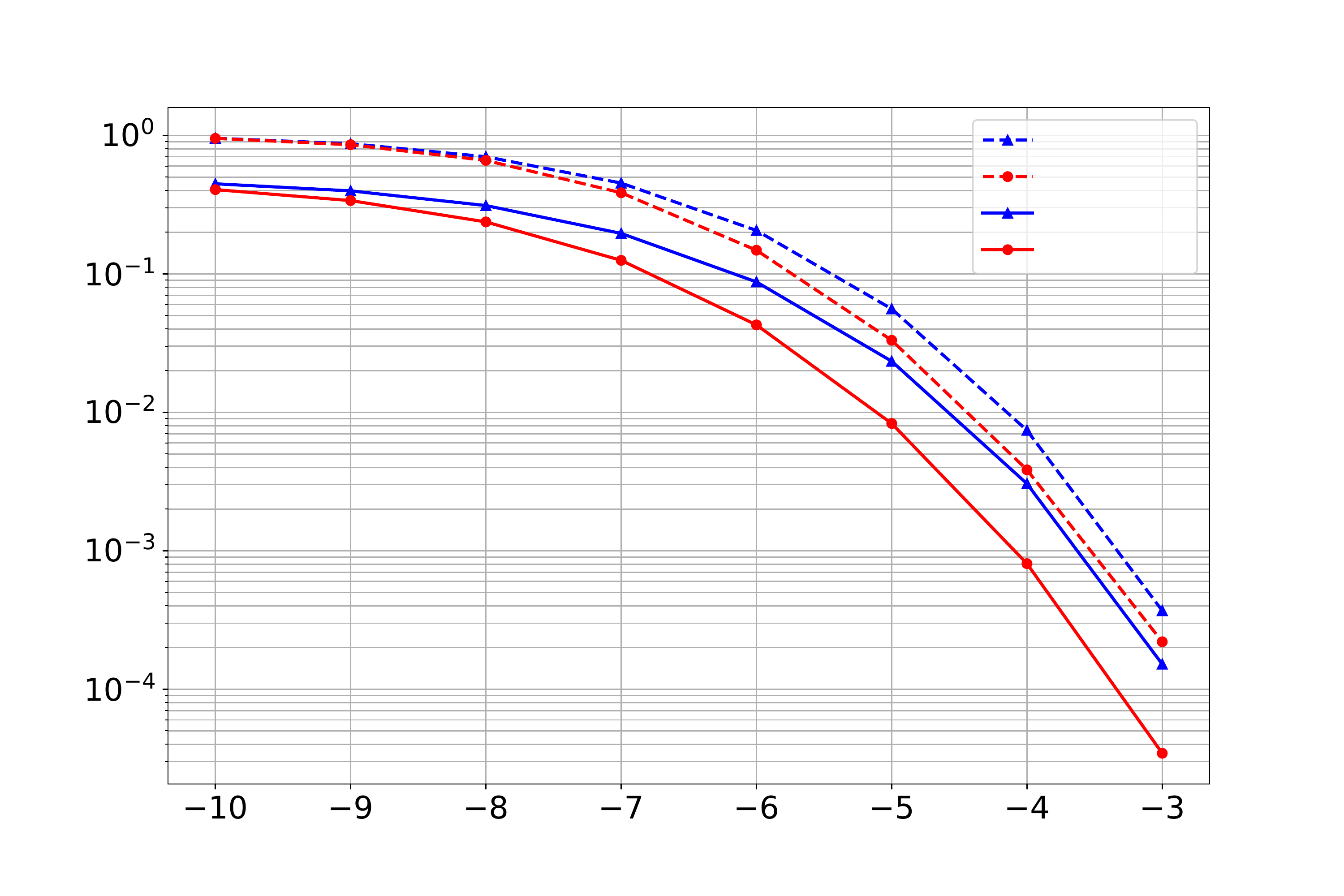}
        \put(-42,108){{\fontsize{5}{6}\selectfont RM BLER}}
    \put(-42,103){{\fontsize{5}{6}\selectfont\KO~BLER}}
    \put(-42,98){{\fontsize{5}{6}\selectfont RM BER}}
    \put(-42,93){{\fontsize{5}{6}\selectfont \KO~BER}}
    \put(-153,0){\footnotesize Signal-to-noise ratio (SNR) [dB]}
    \put(-192,63){\rotatebox[origin=t]{90}{\footnotesize Error rate}}
        \caption{$\KO(8,2)$ trained on AWGN is robust when tested on a bursty channel and maintains a significant gain over $\RM(8,2)$.}
    \label{fig:bursty}
\end{figure}
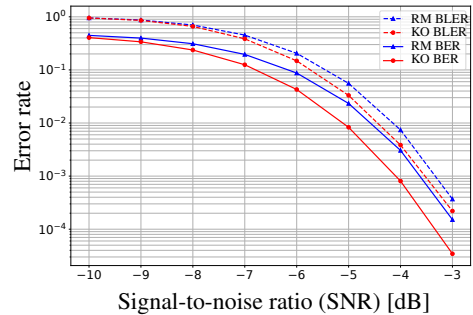

\subsection{Ablation studies}
\label{sec:ablation_studies}



In comparison to the classical RM codes, the KO codes have two additional features: real-valued codewords and non-linearity. It is thus natural to ask how each of these components contribute to its gains over RM codes. To evaluate their contribution, we did ablation experiments for $\KO(8,2)$: (i) First, we constrain the KO codewords to be binary but allow for non-linearity in the encoder $g_\theta$. The performance of this binarized version KO-b$(8,2)$ is illustrated in Figure~\ref{fig:ablation_82} below. We observe that this binarized KO-b$(8,2)$ performs similar to $\RM(8,2)$ except for slight gains at high SNRs but uniformly worse than $\KO(8,2)$. (ii) Now we transmit the real-valued codewords but constrain the encoder $g_\theta$ to be linear (in real-value operations). The resulting code, KO-linear$(8,2)$, performs almost identical to $\RM(8,2)$ but worse than $\KO(8,2)$, as highlighted by the orange curve in Figure~\ref{fig:ablation_82}.

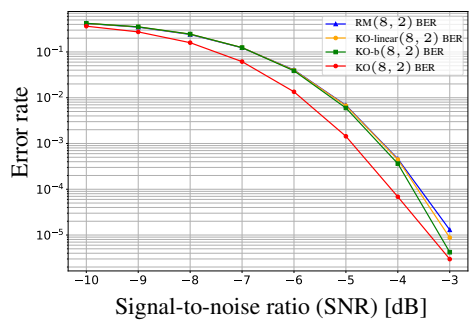
\begin{figure}[H] 
    \centering
    \includegraphics[width=.4\textwidth]{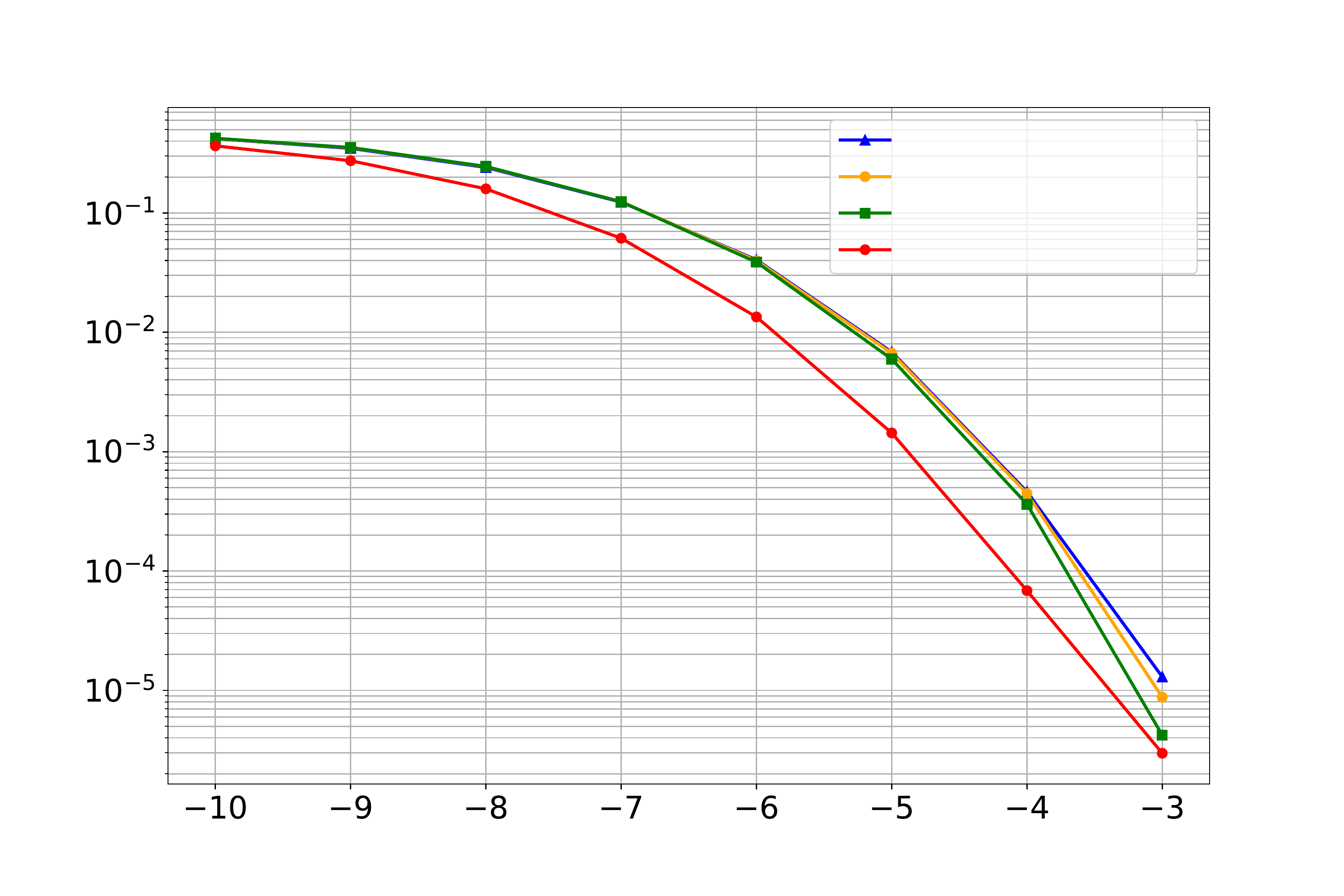}
    \put(-62,109){\fontsize{4}{5}\selectfont RM$(8,2)$ BER}
    \put(-62,103){\fontsize{4}{5}\selectfont \KO-linear$(8,2)$ BER}
    \put(-62,98){\fontsize{4}{5}\selectfont \KO-b$(8,2)$ BER}
    \put(-62,92){\fontsize{4}{5}\selectfont \KO$(8,2)$ BER}
    \put(-153,0){\footnotesize Signal-to-noise ratio (SNR) [dB]}
    \put(-192,63){\rotatebox[origin=t]{90}{\footnotesize Error rate}}
\caption{Ablation studies highlight that both non-linearity and real-valued codewords are equally important for good performance of KO codes. The linear version, KO-linear$(8,2)$, and the binary version, KO-b$(8,2)$, both perform worse than $\KO(8,2)$ and similar to $\RM(8,2)$.}
    \label{fig:ablation_82}
\end{figure}

These ablation experiments suggest us that presence of both the non-linearity and real-valued codewords are necessary for the good performance of KO codes and removal of any of these components hurts the gains it achieves over RM codes. Further, this also highlights that in absence of either of these components, the performance drops back to that of the original RM codes.

\subsection{Complexity of \KO~decoding} 
\label{sec:complexity_kodecode}
Ultra-Reliable Low Latency Communication (URLLC) is increasingly required for modern applications including vehicular communication, virtual reality, and remote robotics \cite{sybis2016channel,jiang2020learn}. 
In general, a \KO$(m,r)$ code requires 
$O(n \log n)$ operations to decode which is the same as the efficient Dumer's decoder for an \RM$(m,r)$ code, where $n=2^m$ is the block length. More precisely, the successive cancellation decoder for \RM$(8,2)$ requires $11268$ operations whereas \KO$(8,2)$ requires $550644$ operations which
we did not try to optimize for this project. 
We discuss promising preliminary results in reducing the computational complexity in \S\ref{sec:tiny_ko}, where KO decoders achieve a  computational efficiency comparable to 
the successive cancellation decoders of RM codes.


\section{KO codes improve upon Polar codes} 
\label{sec:polar}
Results from \S\ref{sec:experimental_results} demonstrate that our KO codes significantly improve upon RM codes on a variety of benchmarks. Here, we focus on a different family of capacity-achieving landmark codes: {\em Polar codes} \citep{arikan2009channel}. 

Polar and RM codes are closely related, especially from an encoding point of view. The generator matrices of both codes are chosen from the same parent square matrix by following different row selection rules. More precisely, consider a RM($m,r$) code that has code dimension $k=\sum_{i=0}^r \binom{m}{i}$ and blocklength $n=2^m$. Its encoding generator matrix is obtained by picking the $k$ rows of the square matrix $\boldsymbol{G}_{n\times n}:=\begin{bmatrix}
0 & 1\\ 1&1
\end{bmatrix}^{\otimes m}$ that have the largest Hamming weights (i.e., Hamming weight of at least $2^{m-r}$), where $[\cdot]^{\otimes m}$ denotes the $m$-th Kronecker power. The Polar encoder, on the other hand, picks the rows of $\boldsymbol{G}_{n\times n}$ that correspond to the most reliable bit-channels \citep{arikan2009channel}.



The recursive Kronecker structure inherent to the parent matrix $\bG_{n \times n}$ can also be represented by a computation graph: a complete binary tree. Thus the corresponding computation tree for a Polar code is obtained by freezing a set of leaves (row-selection). We refer to this encoding computation graph of a Polar code as its { \em Plotkin tree}.
This Plotkin tree structure of Polar codes enables a matching efficient decoder: the {\em successive cancellation} (SC). The SC decoding algorithm is similar to Dumer's decoding for RM codes. Hence, Polar codes can be completely characterized by their corresponding Plotkin trees.



Inspired by the Kronecker structure of Polar Plotkin trees, we design a new family of \KO~codes to strictly improve upon them. We build a novel NN architecture that capitalizes on the Plotkin tree skeleton and generalizes it to nonlinear codes. This enables us to discover new nonlinear algebraic structures. The \KO~encoder and decoder can be trained in an end-to-end manner using variants of stochastic gradient descent (\S\ref{polar_appendix}).

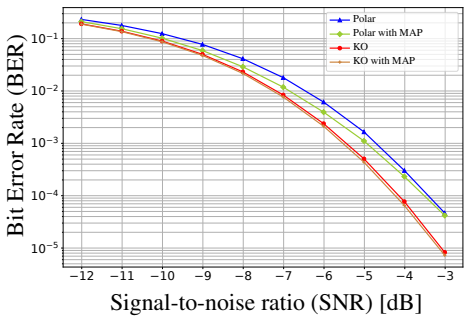
\begin{figure}[H] 
    \centering
    \includegraphics[width=.4\textwidth]{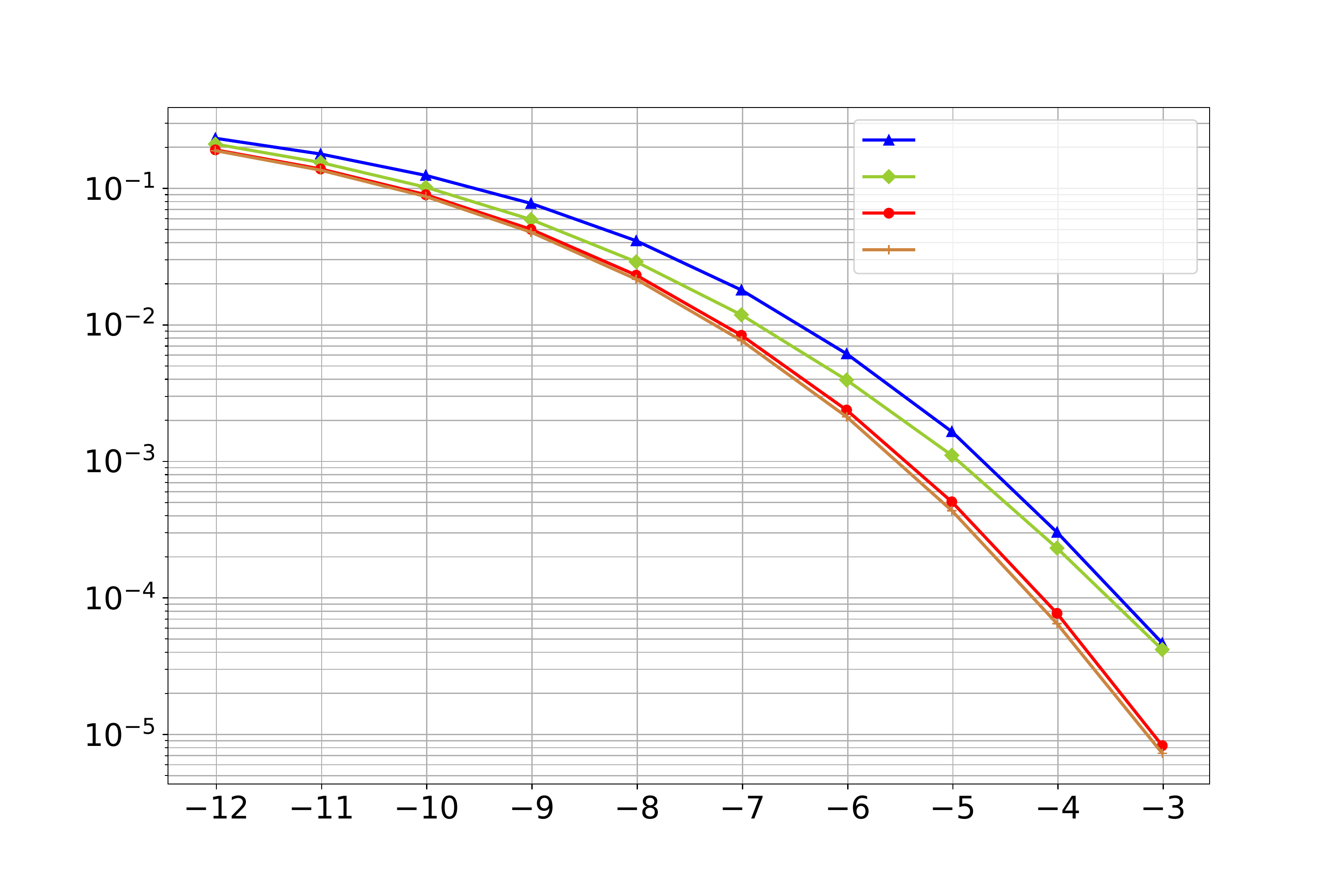}
    \put(-61,109){{\fontsize{4}{5}\selectfont Polar}}
    \put(-61,103.8){\fontsize{4}{5}\selectfont Polar with MAP}
    \put(-61,98.5){\fontsize{4}{5}\selectfont KO}
    \put(-61,93){\fontsize{4}{5}\selectfont KO with MAP}
    \put(-153,0){\footnotesize Signal-to-noise ratio (SNR) [dB]}
    \put(-192,60){\rotatebox[origin=t]{90}{\footnotesize Bit Error Rate (BER)}}
    \vspace{-0.1cm}
\caption{Neural network based KO code improves upon the Polar$(64,7)$ code when trained on AWGN channel. KO decoder also matches the optimal MAP decoder.}
\label{Fig_Polar_BER}
\end{figure}

In Figure~\ref{Fig_Polar_BER}, we compare the performance of our KO code with its competing Polar$(64,7)$ code, i.e., code dimension $k=7$ and block length $n=64$, in terms of BER. Figure~\ref{Fig_Polar_BER} highlights that our KO code achieves significant gains over Polar$(64,7)$ on a wide range of SNRs. In particular, we obtain a gain of almost $0.7$ dB compared to that of Polar at the BER $10^{-4}$. For comparison we also plot the performance of both codes with the optimal MAP decoding. We observe that the BER curve of our KO decoder, unlike the SC decoder, almost matches that of the MAP decoder, convincingly demonstrating its optimality.



We also observe similar improvements for BLER (Figure \ref{Fig_Polar_BLER}, \S\ref{polar_appendix}). This successful case study with training KO (encoder, decoder) pairs further demonstrates that our novel neural architectures seamlessly generalize to codes with an underlying Kronecker product structure.

\section{Tiny KO}
\label{sec:tiny_ko}
In this section we focus on further reducing the total number of mathematical operations required for our KO decoder with the objective of achieving similar computational efficiency as the successive cancellation decoder of RM codes. 

As detailed in \S\ref{sec:architecture}, each neural component in the KO encoder and decoder has $3$ hidden layers with $32$ nodes each. 
For the decoder, the total number of parameters in each decoder neural block is $69\times 32$. We replace all neural blocks with a smaller one with 1 hidden layer of $4$ nodes. 
This decoder neural block has $20$ parameters, obtaining a factor of $110$ compression in the number of parameters. The computational complexity of this compressed decoder, which we refer to as TinyKO, is within a factor of $4$ from  Dumer's successive cancellation decoder. 
Each neural network component has two matrix multiplication steps and one activation function on a vector, which can be fully parallelized on a GPU. 
With the GPU parallelization, TinyKO has the same time complexity/latency as Dumer's SC decoding. 

Table~\ref{tab:tiny} shows that there is almost no loss in reliability for the compressed  $\KO(8,2)$ encoder and decoder in this manner. Training a smaller neural network take about two times more iterations compared to the larger one, although each iteration is faster for the smaller network.  

\begin{table}[h] 
\begin{center}
\begin{tabular}{ |c|r|r| } 
 \hline
 SNR (dB) & TinyKO$(8,2)$ BER  & \KO$(8,2)$ BER\\ \hline
-10 &  0.38414 $\pm$ 2e-7 & 0.36555 $\pm$ 2e-7\\
-9&     0.29671 $\pm$ 2e-7 & 0.27428 $\pm$ 2e-7\\
-8&     0.18037 $\pm$ 2e-7 & 0.15890 $\pm$ 2e-7\\
-7&     0.07455 $\pm$ 2e-7 & 0.06167 $\pm$ 1e-7\\
-6&     0.01797 $\pm$ 8e-8 & 0.01349 $\pm$ 7e-8\\
-5&     2.18083e-3 $\pm$ 3e-8 & 1.46003e-3 $\pm$ 2e-8\\
-4&     1.18919e-4 $\pm$ 7e-9 & 0.64702e-4 $\pm$ 4e-9\\
-3&     4.54054e-6 $\pm$ 1e-9 & 3.16216e-6 $\pm$ 1e-9\\
 \hline
\end{tabular}
\end{center}
\caption{The smaller TinyKO neural architecture with 100 times smaller number of parameters achieve similar bit-error-rates as the bigger KO architecture. }
\label{tab:tiny}
\end{table}

If one is allowed more computation time  (e.g., $O(n^r \log n)$),  then  \cite{ye2020recursive} proposes a recursive projection-aggregation (RPA) decoder for \RM($m,r$)~codes that significantly improves over Dumer's successive cancellation. With list decoding, this is empirically shown to approach the performance of the MAP decoder. It is a promising direction to  explore deep learning architectures upon the computation tree of the RPA decoders to design new family of codes.

\section{Related work} 
\label{sec:related} 

There is tremendous interest in the coding theory community to incorporate deep learning methods. In the context of channel coding, the bulk of the works focus on decoding known linear codes using data-driven neural decoders \citep{nachmani2016learning,  o2017introduction, dorner2017deep, gruber2017deep, nachmani2018deep, kim2018communication, vasic2018learning, teng2019low,  jiang2019deepturbo, nachmani2019hyper, buchberger2020prunin, habib2020learning, chen2021cyclically}; even here, most works have limited themselves to small block lengths due to the difficulty in generalization (for instance, even when nearly $90\%$ of the codewords of a rate $1/2$ Polar code over $8$ information bits are exposed to the neural decoder \citep{gruber2017deep}).  

On the other hand, very few works in the literature focus on discovering {\em both}  encoders and decoders; the few which do, operate at very small block lengths \citep{o2016learning, o2017introduction}. One of the major challenges here is to jointly train the (encoder, decoder) pairs without getting stuck in local optima as the losses are non-convex. In \cite{jiang2019turbo}, the authors employ clever training tricks to learn a  novel autoencoder based codes that outperform the classical Turbo codes, which are sequential in nature. In contrast, here we focus on the generalizations of the Kronecker operation that underpins the RM and Polar family.

 RM and Polar codes have seen active research, especially  on improving decoding  using neural networks: \citep{tallini1995neural, xu2017improved, cammerer2017scaling, bennatan2018deep, doan2018neural,lian2019learned, wang2019learning, carpi2019reinforcement, ebada2019deep}. A common theme across majority of these works is to consider iterative/sequential decoding algorithms, such as belief propagation (BP), bit-flipping (BF), etc., and improve upon their performance by introducing learnable neural network components in them. On the other hand, our KO decoder strictly improves upon the natural SC decoder. Further, we learn the matching KO encoder whereas the encoding is fixed for them. 
 

\section{Conclusion}
\label{sec:conclusion}

We introduce \KO~codes that  generalize the recursive Kronecker operation crucial to designing RM and Polar codes. 
Using the computation tree (known as a Plotkin tree) of these classical codes as a skeleton, we propose a novel neural network architecture tailored for channel  communication. Training over the  AWGN channel, we discover the first  family of {\em non-linear} codes that are not built upon any linear structure. KO codes significantly outperform the baseline Polar and RM codes under similar successive cancellation decoding architectures, which we call Dumer's decoder for the RM codes. The pairwise distance profile reveals that 
\KO~code combines the analytical structure of algebraic codes with the random structure of the celebrated random Gaussian codes.

\section*{Acknowledgements}
Ashok would like to thank his colleagues Mona Zehni and Konik Kothari for helpful discussions about the project.

\bibliographystyle{icml2021}
\bibliography{references_commu,references_consistent}

\clearpage
\onecolumn
\appendix


\section*{Appendix}

\section{Polar$(64,7)$ code}\label{polar_appendix}

Recall from Section~\ref{sec:polar} that the Plotkin tree for a Polar code is obtained by freezing a set of leaves in a complete binary tree. These frozen leaves are chosen according to the reliabilities, or equivalently, error probabilities, of their corresponding bit channels. In other words, we first approximate the error probabilities of all the $n$-bit channels and pick the $k$-smallest of them using the procedure from \cite{tal2013construct}. These $k$ active set of leaves correspond to the transmitted message bits, whereas the remaining $n-k$ frozen leaves always transmit zero.

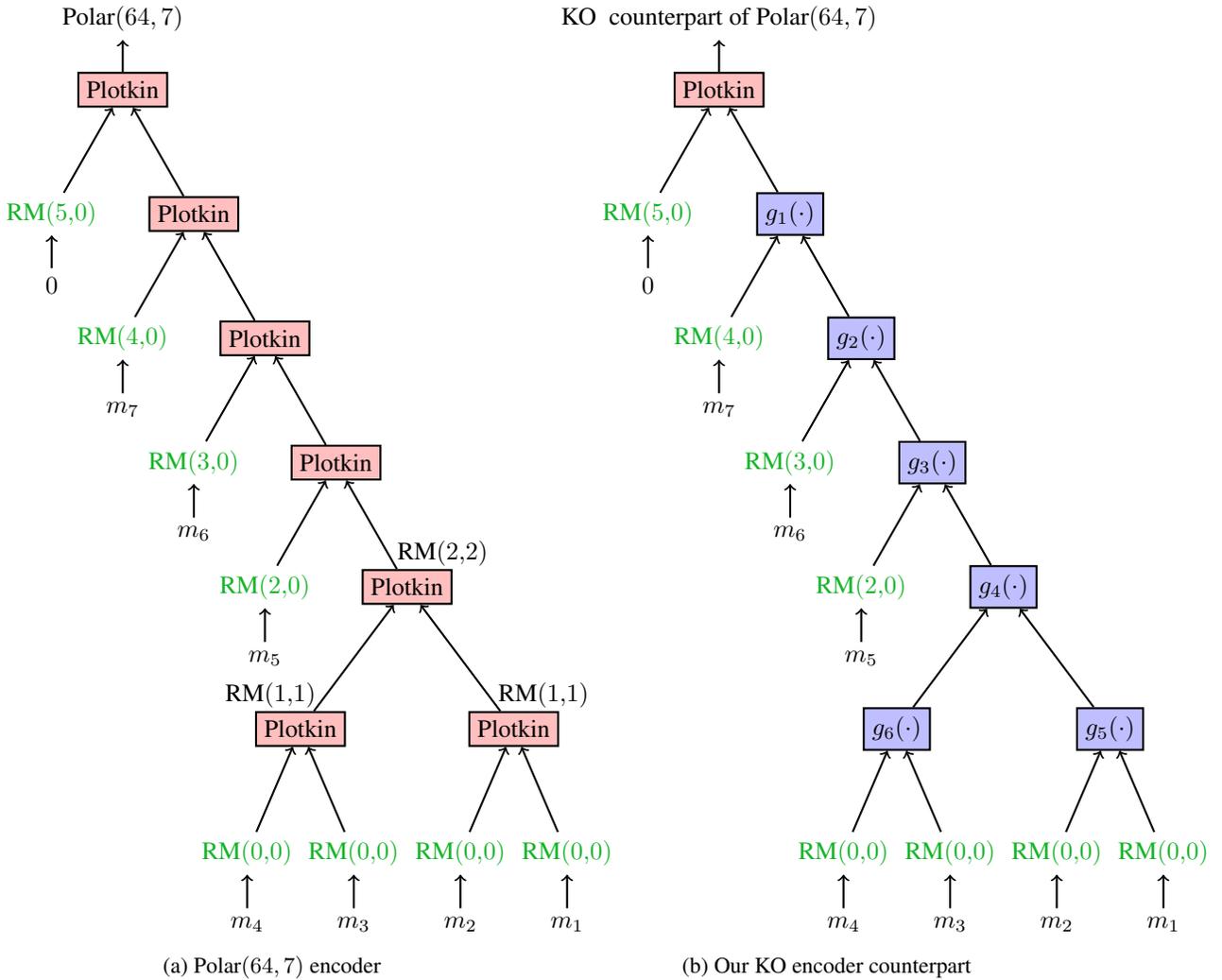
\begin{figure*}[h]
\begin{subfigure}{0.45\textwidth}
	\centering
	\begin{tikzpicture}
	\node at (1,10) (c) {Polar$(64,7)$};
	\node [block, fill=red!25] at (1,9) (r1) {Plotkin};
	\node at (0,7.25) (r2) {\textcolor{MyGreen1}{$\RM(5,\!0)$}};
	\node [block, fill=red!25] at (2,7.25) (r3) {Plotkin};
	\node at (1,5.5) (r4)
	{\textcolor{MyGreen1}{$\RM(4,\!0)$}};
	\node [block, fill=red!25] at (3,5.5) (r5) {Plotkin};
	\node at (2,3.75) (r6)
	{\textcolor{MyGreen1}{$\RM(3,\!0)$}};
	\node [block, fill=red!25] at (4,3.75) (r7) {Plotkin};
	\node at (3,2) (r8)
	{\textcolor{MyGreen1}{$\RM(2,\!0)$}};
	\node [block, fill=red!25] at (5,2) (r9) {Plotkin};
	\node [block, fill=red!25] at (6.5,0) (r10) {Plotkin};
	\node [block, fill=red!25] at (3.5,0) (r11) {Plotkin};
	
	\node at (7.25,-1.75) (r12)
	{\textcolor{MyGreen1}{$\RM(0,\!0)$}};
	\node at (5.75,-1.75) (r13)
	{\textcolor{MyGreen1}{$\RM(0,\!0)$}};
	\node at (4.25,-1.75) (r14)
	{\textcolor{MyGreen1}{$\RM(0,\!0)$}};
	\node at (2.75,-1.75) (r15)
	{\textcolor{MyGreen1}{$\RM(0,\!0)$}};

	\node at (6.92,0.48) (r16) {$\RM(1,\!1)$};
	\node at (3.08,0.48) (r17) {$\RM(1,\!1)$};
	\node at (5.5,2.48) (r18) {$\RM(2,\!2)$};
	
	\node at (0,6.25) (m8) {{$0$}};
	\node at (1,4.5) (m7) {$m_7$};
	\node at (2,2.75) (m6) {$m_6$};
	\node at (3,1) (m5) {$m_5$};
	\node at (2.75,-2.75) (m4) {$m_4$};
	\node at (4.25,-2.75) (m3) {$m_3$};
	\node at (5.75,-2.75) (m2) {$m_2$};
	\node at (7.25,-2.75) (m1) {$m_1$};

	\draw [->,thick] (r1)--(c);
	
	\draw [->,thick] (r2)--(r1);
	\draw [->,thick] (r3)--(r1);
	\draw [->,thick] (r4)--(r3);
	\draw [->,thick] (r5)--(r3);
	\draw [->,thick] (r6)--(r5);
	\draw [->,thick] (r7)--(r5);
	\draw [->,thick] (r8)--(r7);
	\draw [->,thick] (r9)--(r7);
	\draw [->,thick] (r10)--(r9);
	\draw [->,thick] (r11)--(r9);
	\draw [->,thick] (r12)--(r10);
	\draw [->,thick] (r13)--(r10);
	\draw [->,thick] (r14)--(r11);
	\draw [->,thick] (r15)--(r11);

	\draw [->,thick] (m8)--(r2);
	\draw [->,thick] (m7)--(r4);
	\draw [->,thick] (m6)--(r6);
	\draw [->,thick] (m5)--(r8);
	\draw [->,thick] (m4)--(r15);
	\draw [->,thick] (m3)--(r14);
	\draw [->,thick] (m2)--(r13);
	\draw [->,thick] (m1)--(r12);

	\end{tikzpicture}
\caption{Polar$(64,7)$ encoder
	}
	\label{Fig_PolarEnc}
\end{subfigure}
\begin{subfigure}{0.45\textwidth}
	\centering
	\begin{tikzpicture}
	\node at (1,10) (c) {\KO~ counterpart of Polar$(64,7)$};
	\node [block, fill=red!25] at (1,9) (r1) {Plotkin};
	\node at (0,7.25) (r2) {\textcolor{MyGreen1}{$\RM(5,\!0)$}};
	\node [block, fill=blue!25] at (2,7.25) (r3) {$g_1(\cdot)$};
	\node at (1,5.5) (r4)
	{\textcolor{MyGreen1}{$\RM(4,\!0)$}};
	\node [block, fill=blue!25] at (3,5.5) (r5) {$g_2(\cdot)$};
	\node at (2,3.75) (r6)
	{\textcolor{MyGreen1}{$\RM(3,\!0)$}};
	\node [block, fill=blue!25] at (4,3.75) (r7) {$g_3(\cdot)$};
	\node at (3,2) (r8)
	{\textcolor{MyGreen1}{$\RM(2,\!0)$}};
	\node [block, fill=blue!25] at (5,2) (r9) {$g_4(\cdot)$};
	\node [block, fill=blue!25] at (6.5,0) (r10) {$g_5(\cdot)$};
	\node [block, fill=blue!25] at (3.5,0) (r11) {$g_6(\cdot)$};
	
	\node at (7.25,-1.75) (r12)
	{\textcolor{MyGreen1}{$\RM(0,\!0)$}};
	\node at (5.75,-1.75) (r13)
	{\textcolor{MyGreen1}{$\RM(0,\!0)$}};
	\node at (4.25,-1.75) (r14)
	{\textcolor{MyGreen1}{$\RM(0,\!0)$}};
	\node at (2.75,-1.75) (r15)
	{\textcolor{MyGreen1}{$\RM(0,\!0)$}};

	
	\node at (0,6.25) (m8) {{$0$}};
	\node at (1,4.5) (m7) {$m_7$};
	\node at (2,2.75) (m6) {$m_6$};
	\node at (3,1) (m5) {$m_5$};
	\node at (2.75,-2.75) (m4) {$m_4$};
	\node at (4.25,-2.75) (m3) {$m_3$};
	\node at (5.75,-2.75) (m2) {$m_2$};
	\node at (7.25,-2.75) (m1) {$m_1$};

	\draw [->,thick] (r1)--(c);
	
	\draw [->,thick] (r2)--(r1);
	\draw [->,thick] (r3)--(r1);
	\draw [->,thick] (r4)--(r3);
	\draw [->,thick] (r5)--(r3);
	\draw [->,thick] (r6)--(r5);
	\draw [->,thick] (r7)--(r5);
	\draw [->,thick] (r8)--(r7);
	\draw [->,thick] (r9)--(r7);
	\draw [->,thick] (r10)--(r9);
	\draw [->,thick] (r11)--(r9);
	\draw [->,thick] (r12)--(r10);
	\draw [->,thick] (r13)--(r10);
	\draw [->,thick] (r14)--(r11);
	\draw [->,thick] (r15)--(r11);

	\draw [->,thick] (m8)--(r2);
	\draw [->,thick] (m7)--(r4);
	\draw [->,thick] (m6)--(r6);
	\draw [->,thick] (m5)--(r8);
	\draw [->,thick] (m4)--(r15);
	\draw [->,thick] (m3)--(r14);
	\draw [->,thick] (m2)--(r13);
	\draw [->,thick] (m1)--(r12);

	\end{tikzpicture}
	\caption{Our \KO~encoder counterpart
	}
	\label{Fig_PolarKOEnc}
\end{subfigure}
\caption{Plotkin trees for the Polar$(64,7)$ encoder and our neural \KO~encoder counterpart. Both codes have dimension $k=7$ and blocklength $n=64$.}
\label{Fig_Polarm6k7}
\end{figure*}

Here we focus on a specific Polar code: Polar$(64,7)$, with code dimension $k=7$ and blocklength $n=64$. For Polar$(64,7)$, we obtain these active set of leaves to be $\mathcal{A}=\{48,56,60,61,62,63,64\}$, and the frozen set to be $\mathcal{A}^{c}=\{1,2,\cdots 64\}\setminus\mathcal{A}$. Using these set of indices and simplifying the redundant branches, we obtain the Plotkin tree for Polar$(64,7)$ to be Figure~\ref{Fig_Polarm6k7}. We observe that this Polar Plotkin tree shares some similarities with that of a $\RM(6,1)$ code (with same $k=7$ and $n=64$) with key differences at the topmost and bottom most leaves.

Capitalizing on the encoding tree structure of Polar$(64,7)$, we build a corresponding \KO~encoder $g_\theta$ which inherits this tree skeleton. In other words, we generalize the Plotkin mapping blocks at the internal nodes of tree, except for the root node, and replace them with a corresponding neural network $g_i$. Figure~\ref{Fig_Polarm6k7} depicts the Plotkin tree of our KO encoder. The \KO~decoder $f_\phi$ is designed similarly. Training of the (encoder, decoder) pair $(g_\theta, f_\phi)$ is similar to that of the $\KO(8,2)$ training which we detail in \S\ref{sec:experimental_results}.

\begin{figure}[H] 
    \centering
    \includegraphics[width=.7\textwidth]{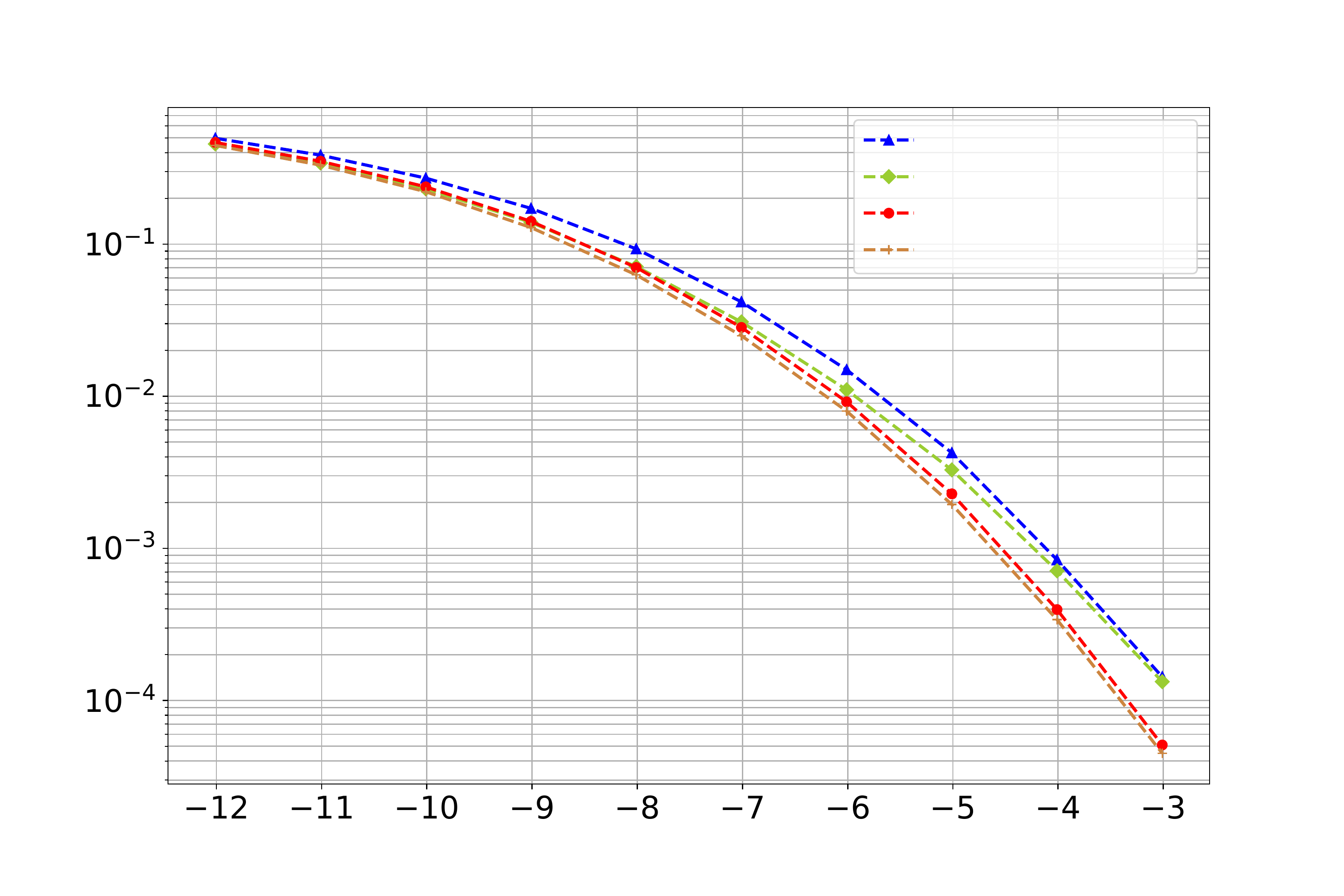}
    \put(-106,191){{\fontsize{7}{7}\selectfont Polar}}
    \put(-106,181.7){\fontsize{7}{7}\selectfont Polar with MAP}
    \put(-106,172.2){\fontsize{7}{7}\selectfont KO}
    \put(-106,162.7){\fontsize{7}{7}\selectfont KO with MAP}
    \put(-223,7){\footnotesize Signal-to-noise ratio (SNR) [dB]}
    \put(-330,110){\rotatebox[origin=t]{90}{\footnotesize Block error rate (BLER)}}
\caption{KO code achieves a significant gain over the Polar$(64,7)$ code in BLER when trained on AWGN channel. KO decoder also matches the optimal MAP decoder.}
\label{Fig_Polar_BLER}
\end{figure}
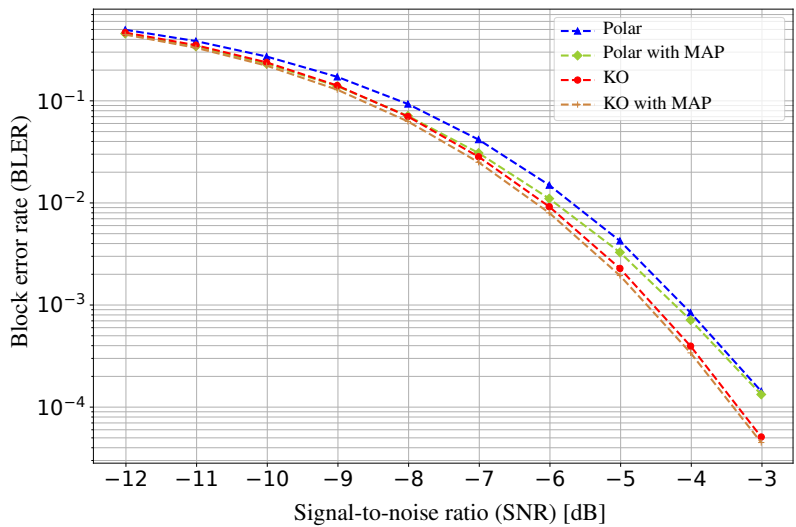

Figure~\ref{Fig_Polar_BLER} shows the BLER performance of the Polar$(64,7)$ code and its competing KO code, for the AWGN channel. Similar to the BER performance analyzed in Figure \ref{Fig_Polar_BER}, the KO code is able to significantly improve the BLER performance. For example, we achieve a gain of around $0.5$ dB when KO encoder is combined with the MAP decoding. Additionally, the close performance of the KO decoder to that of the MAP decoder confirms its optimality.

\section{Gains with list decoding}
\label{sec:gains_list}

Successive cancellation decoding  can  be significantly improved by list decoding. List decoding allows one to gracefully tradeoff 
computational complexity and reliability by maintaining a list (of a fixed size) of candidate codewords during the  decoding process. 
The following figure demonstrates that 
\KO(8,2) code with list decoding enjoys a significant  gain over the non-list counterpart. 
This promising result opens several interesting directions, which  are current focuses of active research. 

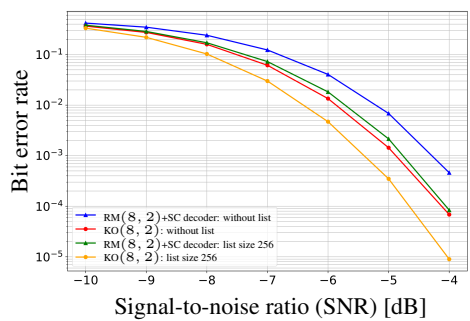
\begin{figure}[h]
    \centering
    \includegraphics[width=.4\textwidth]{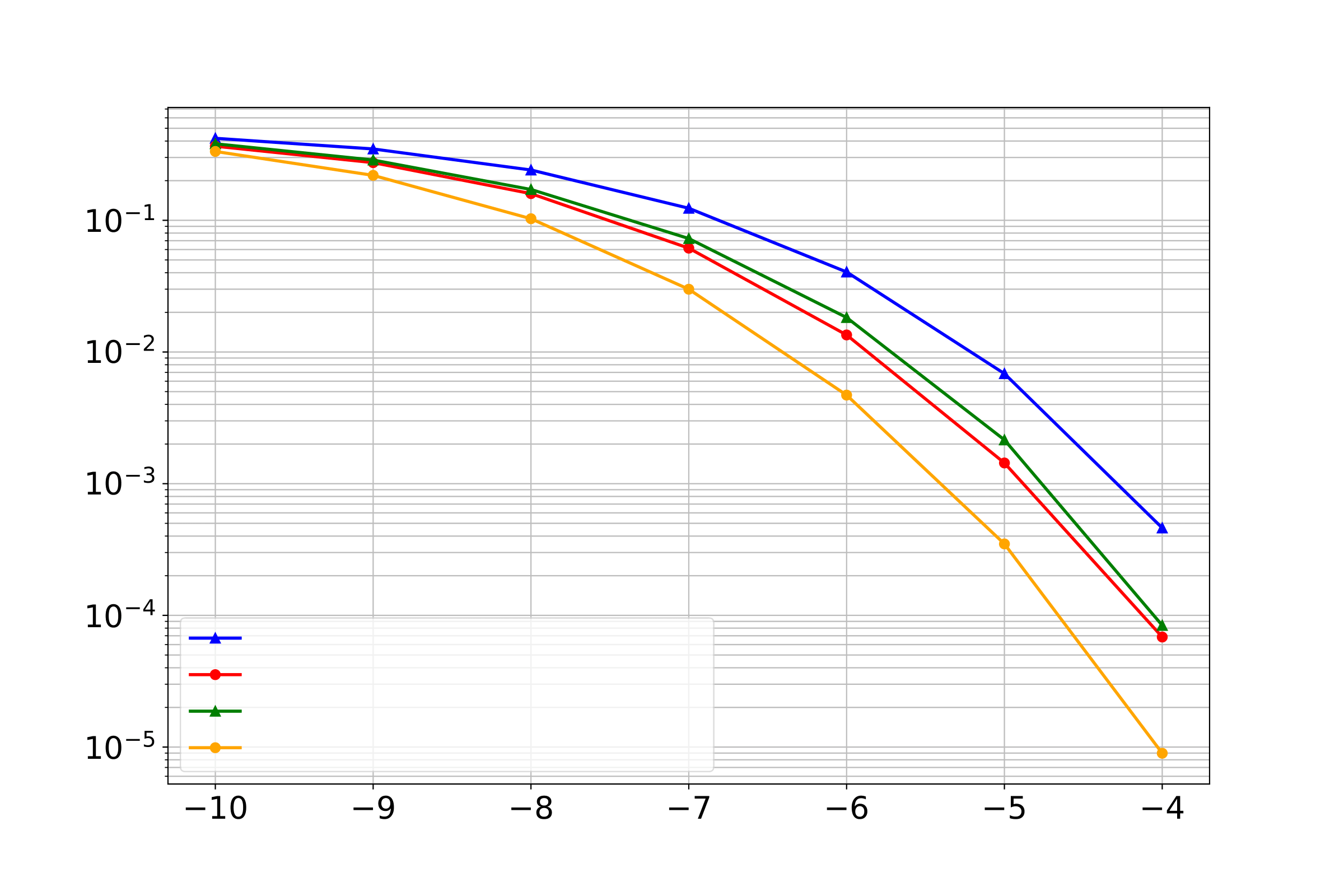}
    \put(-157,35){\fontsize{4}{5}\selectfont RM$(8,2)$+SC decoder: without list}
    \put(-157,30){\fontsize{4}{5}\selectfont \KO$(8,2)$: without list}
    \put(-157,25){\fontsize{4}{5}\selectfont RM$(8,2)$+SC decoder: list size 256}
    \put(-157,20){\fontsize{4}{5}\selectfont \KO$(8,2)$: list size 256}
    \put(-153,0){\footnotesize Signal-to-noise ratio (SNR) [dB]}
    \put(-192,63){\rotatebox[origin=t]{90}{\footnotesize Bit error rate}}
    \vspace{-0.1cm}
    \caption{The same \KO(8,2) encoder and decoder as those used in Figure~\ref{fig:awgn82} achieve a significant gain (without any retraining or fine-tuning) when list decoding is used together with the \KO~decoder. The magnitude of the gain is comparable to the gain achieved by the same list decoding technique on  the  successive cancellation decoder of the \RM(8,2) code. We used the list decoding  from \cite{dumer2006soft} but without the permutation technique. }
    \label{fig:conclusion}
\end{figure}

Polar codes with list decoding achieves the state-of-the-art performances   \cite{tal2015list}. 
It is a promising direction to design   large block-lengths \KO~codes (based on the skeleton of Polar codes) that  can  improve upon the state-of-the-art  list-decoded Polar codes. One direction is to train \KO~codes as we propose and include list decoding after the training. 
A more ambitious direction is to include list decoding in the training, potentially further improving the performance by discovering an encoder tailored for list decoding.

Unlike Polar codes,  RM~codes have an extra structure of an algebraic symmetry;   a \RM~codebook is invariant under certain permutations.  
This can be exploited in list decoding as shown in  \cite{dumer2006soft}, to get a further gain over what is shown in Figure~\ref{fig:conclusion}.
However, when a \KO~code is trained based on a \RM~skeleton, this symmetry is lost. A question of interest is whether one can discover nonlinear codes with such symmetry.

\section{Discussion} 
\label{sec:ablation}

\subsection{On modulation and practicality of KO codes}
\label{sec:modulation}
We note that our real-valued KO codewords are {\em entirely practical} in wireless communication; the peak energy of a symbol  is only $22.48\%$ larger than the average in KO codes. The impact on the power amplifier is not any different from that of a more traditional modulation (like $16$-QAM). 
Training KO code is a form of {\em jointly} designing the coding and modulation steps; this approach has a long history in wireless communication (e.g., Trellis coded modulation) but the performance gains have been restricted by the human ingenuity in constructing the heuristics.
 
\subsection{Comparison with LDPC and BCH codes}
\label{sec:ldpc_bch}
We expect good performance for BCH at the short blocklengths considered in the paper though with high-complexity (polynomial time) decoders such as ordered statistics decoder (OSD). On the other hand, there does not exist good LDPC codes at the $k$ and $n$ regimes of this paper; thus, we do not expect good performance for LDPC at these regimes.

\section{Plotkin construction}
\label{sec:app_plotkin}

\citet{plotkin60} proposed this scheme in order to combine two codes of smaller code lengths and construct a larger code with the following properties. 
It is relatively easy to construct a code with either a high rate but a small distance (such as sending the raw information bits directly) or a large distance but a low rate (such as repeating each bit multiple  times).
Plotkin construction combines 
such two codes of rates $\rho_{\boldsymbol{u}} > \rho_{\boldsymbol{v}}$ 
and distances $d_{\boldsymbol{u}}<d_{\boldsymbol{v}}$, to design a larger block length code satisfying rate $\rho = (\rho_{\boldsymbol{u}} + \rho_{\boldsymbol{v}})/2$ and distance  $\min\{2d_{\boldsymbol{u}}, d_{\boldsymbol{v}}\}$. This significantly improves upon a simple time-sharing of those codes, which achieves the same rate but  distance only $\min\{d_{\boldsymbol{u}}, d_{\boldsymbol{v}}\}$.

{\bf Note:} Following the standard convention, we fix the leaves in the Plotkin tree of a first order $\RM(m,1)$ code  to be zeroth order RM codes and the full-rate $\RM(1,1)$ code. On the other hand, a second order $\RM(m,2)$ code contains the first order RM codes and the full-rate $\RM(2,2)$ as its leaves.


\section{$\KO(8,2)$: Architecture and training}
\label{sec:rm_82}

As highlighted in \S\ref{sec:experimental_results}, our KO codes improve upon RM codes significantly on a variety of benchmarks. We present the architectures of the $\KO(8,2)$ encoder and the $\KO(8,2)$ decoder, and their joint training methodology that are crucial for this superior performance.

\begin{figure*}[h]
\centering
\begin{subfigure}{0.46\textwidth}
	\centering
	\begin{tikzpicture}
	\node at (1,10) (c) {$\RM(8,\!2)$};
	\node [block, fill=red!25] at (1,9) (r1) {Plotkin};
	\node at (-0.25,7.25) (r2) {\textcolor{MyGreen1}{$\RM(7,\!1)$}};
	\node [block, fill=red!25] at (2.25,7.25) (r3) {Plotkin};
	\node at (1,5.5) (r4)
	{\textcolor{MyGreen1}{$\RM(6,\!1)$}};
	\node at (3.5,5.5) (r5) {{$\RM(6,\!2)$}};
	\node [block, fill=red!25] at (4.75,3.75) (r6) {Plotkin};
	\node at (3.5,2) (r7)
	{\textcolor{MyGreen1}{$\RM(2,\!1)$}};
	\node at (6,2) (r8) {\textcolor{MyGreen1}{$\RM(2,\!2)$}};
	
	\node at (2.75,7.75) (r9) {{$\RM(7,\!2)$}};
	\node at (5.25,4.25) (r10) {{$\RM(3,\!2)$}};
	
	\node at (-0.25,6.25) (m7) {{$\bma_{(7,1)}$}};
	\node at (1,4.5) (m6) {$\bma_{(6,1)}$};
	\node at (3.5,1) (m21) {$\bma_{(2,1)}$};
	\node at (6,1) (m22) {$\bma_{(2,2)}$};

	\draw [->,thick] (r1)--(c);
	
	\draw [->,thick] (r2)--(r1);
	\draw [->,thick] (r3)--(r1);
	\draw [->,thick] (r4)--(r3);
	\draw [->,thick] (r5)--(r3);
	\draw [., dashed, thick] (r6)--(r5);
	\draw [->,thick] (r7)--(r6);
	\draw [->,thick] (r8)--(r6);
	
	\draw [->,thick] (m7)--(r2);
	\draw [->,thick] (m6)--(r4);
	\draw [->,thick] (m21)--(r7);
	\draw [->,thick] (m22)--(r8);
	\end{tikzpicture}
	\caption{$\RM(8,\!2)$ encoder
	}
	\label{Fig_RMEnc82}
\end{subfigure}
\hfill
\centering
\begin{subfigure}{0.46\textwidth}
	\centering
	\begin{tikzpicture}
	\node at (1,10) (c) {$\KO(8,\!2)$};
	\node [block, fill=blue!25] at (1,9) (r1) {$g_1(\cdot)$};
	\node at (-0.25,7.25) (r2) {\textcolor{MyGreen1}{$\RM(7,\!1)$}};
	\node [block, fill=blue!25] at (2.25,7.25) (r3) {$g_2(\cdot)$};
	\node at (1,5.5) (r4)
	{\textcolor{MyGreen1}{$\RM(6,\!1)$}};
	\node at (3.5,5.5) (r5) {};
	\node [block, fill=blue!25] at (4.75,3.75) (r6) {$g_6(\cdot)$};
	\node at (3.5,2) (r7)
	{\textcolor{MyGreen1}{$\RM(2,\!1)$}};
	\node at (6,2) (r8) {\textcolor{MyGreen1}{$\RM(2,\!2)$}};
	
	
	\node at (-0.25,6.25) (m7) {{$\bma_{(7,1)}$}};
	\node at (1,4.5) (m6) {$\bma_{(6,1)}$};
	\node at (3.5,1) (m21) {$\bma_{(2,1)}$};
	\node at (6,1) (m22) {$\bma_{(2,2)}$};

	\draw [->,thick] (r1)--(c);
	
	\draw [->,thick] (r2)--(r1);
	\draw [->,thick] (r3)--(r1);
	\draw [->,thick] (r4)--(r3);
	\draw [->,thick] (r5)--(r3);
	\draw [., dashed, thick] (r6)--(r5);
	\draw [->,thick] (r7)--(r6);
	\draw [->,thick] (r8)--(r6);
	
	\draw [->,thick] (m7)--(r2);
	\draw [->,thick] (m6)--(r4);
	\draw [->,thick] (m21)--(r7);
	\draw [->,thick] (m22)--(r8);
	\end{tikzpicture}
	\caption{$\KO(8,\!2)$ encoder
	}
	\label{Fig_NEnc82}
\end{subfigure}
\caption{Plotkin trees for $\RM(8,\!2)$ and $\KO(8,\!2)$ encoders. Leaves are highlighted in green. Both codes have dimension $k=37$ and blocklength $n=256$.}
\end{figure*}
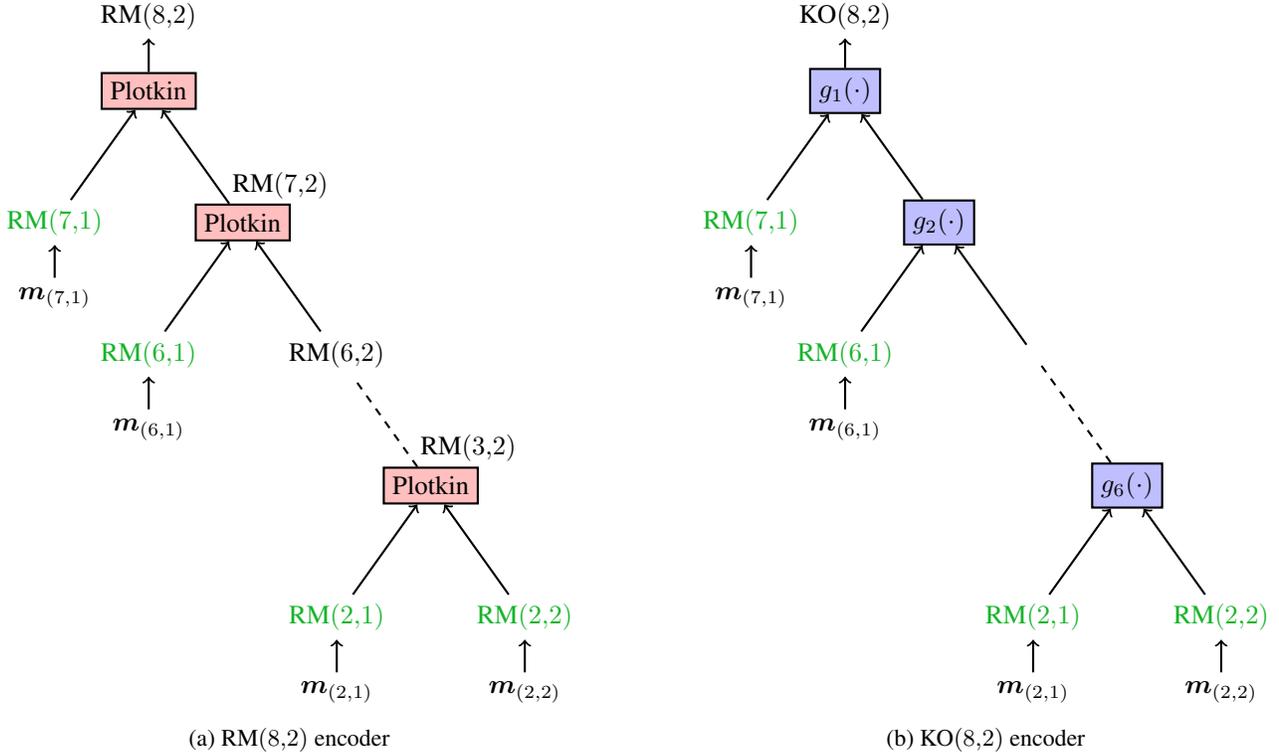

\subsection{$\KO(8,2)$ encoder}
\label{sec:enc82}
{\bf Architecture.} $\KO(8,2)$ encoder inherits the same Plotkin tree structure as that of the second order $\RM(8,2)$ code and thus RM codes of first order and the second order $\RM(2,2)$ code constitute the leaves of this tree, as highlighted in Figure~\ref{Fig_NEnc82}. On the other hand, a critical distinguishing component of our $\KO(8,2)$ encoder is a set of encoding neural networks $g_\theta = \{ g_1, \ldots, g_6 \}$ that strictly generalize the Plotkin mapping. In other words, we associate a neural network $g_i \in g_\theta$ to each internal node $i$ of this tree. If $\bv$ and $\bu$ denote the codewords arriving from left and right branches at this node, we combine them non-linearly via the operation $(\bu, \bv) \mapsto g_i(\bu, \bv)$. 

We carefully parametrize each encoding neural network $g_i$ so that they generalize the classical Plotkin map $\mathrm{Plotkin}(\bu,\bv)= (\bu, \bu \oplus \bv)$. In particular, we represent them as $g_i(\bu, \bv) = (\bu, \tilde{g}_i(\bu, \bv) + \bu \oplus \bv ) $, where $\tilde{g}_i:\reals^2 \to \reals$ is a neural network of input dimension $2$ and output size $1$. Here $\tilde{g_i}$ is applied coordinate-wise on its inputs $\bu$ and $\bv$. This clever parametrization can also be viewed as a skip connection on top of the Plotkin map. Similar skip-like ideas have been successfully used in the literature though in a different context of learning decoders \cite{neural_augment}. On the other hand, we exploit these ideas for both encoders and decoders which further contribute to significant gains over RM codes.



{\bf Encoding.} From an encoding perspective, recall that the $\KO(8,2)$ code has code dimension $k=37$ and block length $n=256$. Suppose we wish to transmit a set of $37$ message bits denoted as $\bma=(\bma_{(2,2)}, \bma_{(2,1)}, \ldots, \bma_{(7,1)})$ through our $\KO(8,2)$ encoder. We first encode the block of four message bits $\bma_{(2,2)}$ into a $\RM(2,2)$ codeword $\bc_{(2,2)}$ using its corresponding encoder at the bottom most leaf of the Plotkin tree. Similarly we encode the next three message bits $\bma_{(2,1)}$ into an $\RM(2,1)$ codeword $\bc_{(2,1)}$. We combine these codewords using the neural network $g_6$ at their parent node, which yields the codeword $\bc_{(3,2)} = g_6(\bc_{(2,2)}, \bc_{(2,1)}  )  \in \reals^8$. The codeword $\bc_{(3,2)}$ is similarly combined with its corresponding left codeword and this procedure is thus recursively carried out till we reach the top most node of the tree, which outputs the codeword $\bc_{(8,2)} \in \reals^{256}$. Finally we obtain the unit-norm $\KO(8,2)$ codeword $\bx$ by normalizing $\bc_{(8,2)}$, \ie $\bx = \bc_{(8,2)}/\norm{\bc_{(8,2)}}_2$.

Note that the map of encoding the message bits $\bma$ into the codeword $\bx$, \ie $\bx = g_\theta(\bma)$, is differentiable with respect to $\theta$ since all the underlying operations at each node of the Plotkin tree are differentiable.

\subsection{$\KO(8,2)$ decoder}
\label{sec:dec82}
\begin{figure*}[h]
\centering
\begin{subfigure}{0.48\textwidth}
	\centering
	\begin{tikzpicture}
	\node at (-0.2,9) (y) {$\boldsymbol{y}$};
	\node [block, fill=red!25] at (1,9) (r1) {$\boldsymbol{L}$};
	\node [block, fill=MyGreen1!25] at (-0.25,7.25) (r2) {{{MAP dec.}}};
	\node [block, fill=red!25] at (2.25,7.25) (r3) {$\boldsymbol{L}_{\boldsymbol{u}}$};
	\node [block, fill=MyGreen1!25] at (1,5.5) (r4) {{{MAP dec.}}};
	\node at (3.5,5.5) (r5) {};
	\node [block, fill=red!25] at (4.75,3.75) (r6) {$\boldsymbol{L}_{\boldsymbol{u}}^{(3,2)}$};
	\node [block, fill=MyGreen1!25] at (3.5,2) (r7) {{{MAP dec.}}};
	\node [block, fill=MyGreen1!25] at (6,2) (r8) {{{MAP dec.}}};
	
	\node at (-0.25,6.25) (m7) {{$\boldsymbol{\hat{m}}_{(7,1)}$}};
	\node at (1,4.5) (m6) {$\boldsymbol{\hat{m}}_{(6,1)}$};
	\node at (3.5,1) (m21) {$\boldsymbol{\hat{m}}_{(2,1)}$};
	\node at (6,1) (m22) {$\boldsymbol{\hat{m}}_{(2,2)}$};

	\node at (0, 8.25) (f3) {{$\mathrm{LSE}$}};
	\node at (1.9, 8.25) (f4) {{$\oplus_{\hat{\bv}}$}};
	\node at (1.25, 6.5) (f3) {{$\mathrm{LSE}$}};
	\node at (3.15, 6.5) (f4) {{$\oplus_{\hat{\bv}}$}};
	\node at (3.75, 3) (f3) {{$\mathrm{LSE}$}};
	\node at (5.65, 3) (f4) {{$\oplus_{\hat{\bv}}$}};

	\draw [->,thick] (y)--(r1);
	
	\draw [->,thick] (r1)--(r2);
	\draw [->,thick] (r1)--(r3);
	\draw [->,thick] (r3)--(r4);
	\draw [->,thick] (r3)--(r5);
	\draw [., dashed, thick] (r6)--(r5);
	\draw [->,thick] (r6)--(r7);
	\draw [->,thick] (r6)--(r8);
	
	\draw [->,thick] (r2)--(m7);
	\draw [->,thick] (r4)--(m6);
	\draw [->,thick] (r7)--(m21);
	\draw [->,thick] (r8)--(m22);
	
	\draw [->,thick] (0.25,6.25) to [out=0,in=-90] (r1.south);
	\draw [->,thick] (1.5,4.5) to [out=0,in=-90] (r3.south);
	\draw [->,thick] (4,1) to [out=0,in=-90] (r6.south);
	
	\draw [->,densely dashed,thick,red] (-0.25,5.95) to [out=-90,in=180] (m6.west);
	\draw [->,densely dashed,thick,red] (1,4.2) to [out=-90,in=180] (m21.west);
	\draw [->,densely dashed,thick,red] (3.5,0.7) to [out=-90,in=-90] (m22.south);

	\end{tikzpicture}
	\caption{$\RM(8, 2)$ decoder
	}
	\label{Fig_RMDec82}
\end{subfigure}
\hfill
\centering
\begin{subfigure}{0.48\textwidth}
	\centering
	\begin{tikzpicture}
	\node [block, fill=blue!25] at (1,9) (r1) {$\boldsymbol{y}$};
	\node [block, fill=MyGreen1!25] at (-0.25,7.25) (r2) {{{Soft-MAP}}};
	\node [block, fill=blue!25] at (2.25,7.25) (r3) {$\boldsymbol{y}_{\boldsymbol{u}}$};
	\node [block, fill=MyGreen1!25] at (1,5.5) (r4) {{{Soft-MAP}}};
	\node at (3.5,5.5) (r5) {};
	\node [block, fill=blue!25] at (4.75,3.75) (r6) {$\boldsymbol{y}_{\boldsymbol{u}}^{(3,2)}$};
	\node [block, fill=MyGreen1!25] at (3.5,2) (r7) {{{Soft-MAP}}};
	\node [block, fill=MyGreen1!25] at (6,2) (r8) {Soft-MAP};
	
	\node at (-0.25,6.25) (m7) {{$\boldsymbol{\hat{m}}_{(7,1)}$}};
	\node at (1,4.5) (m6) {$\boldsymbol{\hat{m}}_{(6,1)}$};
	\node at (3.5,1) (m21) {$\boldsymbol{\hat{m}}_{(2,1)}$};
	\node at (6,1) (m22) {$\boldsymbol{\hat{m}}_{(2,2)}$};

	\node at (0.17, 8.25) (f3) {{$f_1$}};
	\node at (1.9, 8.25) (f4) {{$f_2$}};
	\node at (1.4, 6.5) (f3) {{$f_3$}};
	\node at (3.15, 6.5) (f4) {{$f_4$}};
	\node at (3.83, 3) (f3) {{$f_{11}$}};
	\node at (5.7, 3) (f4) {{$f_{12}$}};
	
	\draw [->,thick] (r1)--(r2);
	\draw [->,thick] (r1)--(r3);
	\draw [->,thick] (r3)--(r4);
	\draw [->,thick] (r3)--(r5);
	\draw [., dashed, thick] (r6)--(r5);
	\draw [->,thick] (r6)--(r7);
	\draw [->,thick] (r6)--(r8);
	
	\draw [->,thick] (r2)--(m7);
	\draw [->,thick] (r4)--(m6);
	\draw [->,thick] (r7)--(m21);
	\draw [->,thick] (r8)--(m22);
	
	\draw [->,thick] (0.25,6.25) to [out=0,in=-90] (r1.south);
	\draw [->,thick] (1.5,4.5) to [out=0,in=-90] (r3.south);
	\draw [->,thick] (4,1) to [out=0,in=-90] (r6.south);
	
	\draw [->,densely dashed,thick,red] (-0.25,5.95) to [out=-90,in=180] (m6.west);
	\draw [->,densely dashed,thick,red] (1,4.2) to [out=-90,in=180] (m21.west);
	\draw [->,densely dashed,thick,red] (3.5,0.7) to [out=-90,in=-90] (m22.south);

	\end{tikzpicture}
	\caption{$\KO(8,2)$ decoder
	}
	\label{Fig_NDec82}
\end{subfigure}
\caption{Plotkin tres for the $\RM(8,2)$ and $\KO(8, 2)$ decoders. Red arrows indicate the bit decoding order.}
\label{Fig_RM82_Decoders}
\end{figure*}
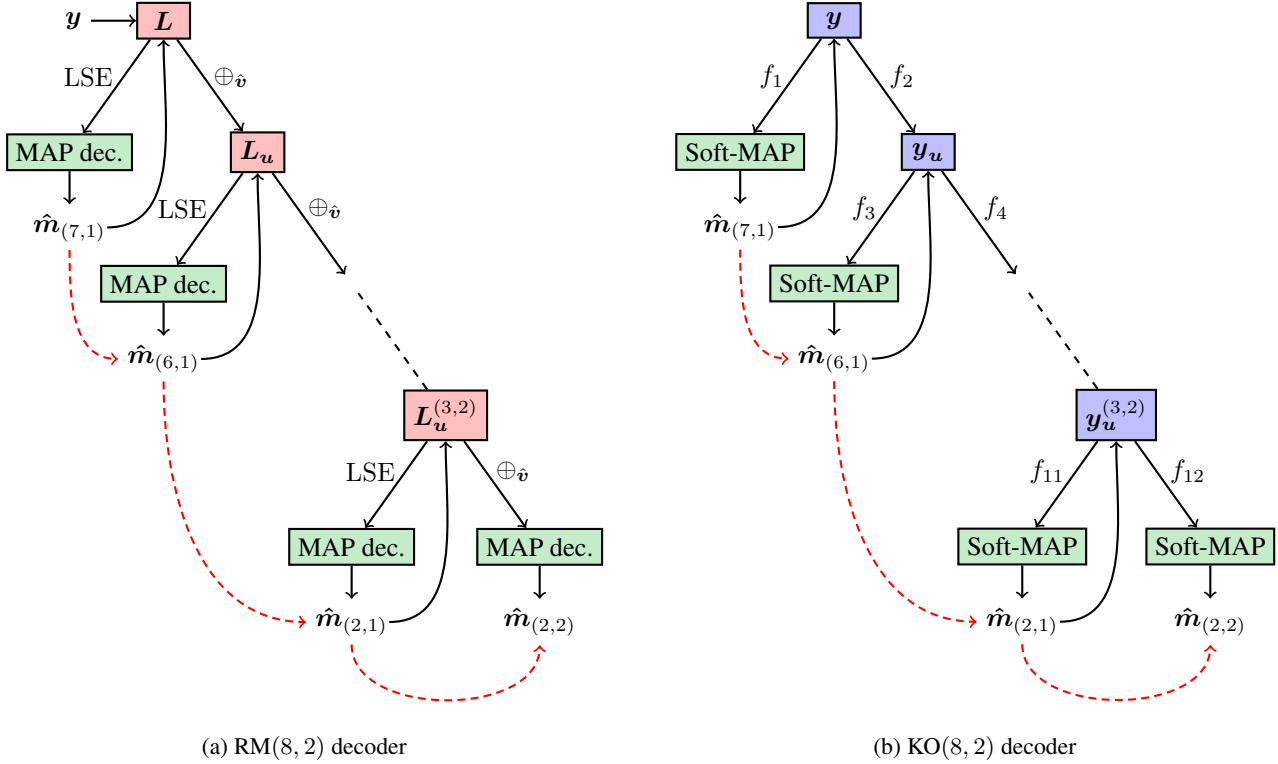

{\bf Architecture.} Capitalizing on the recursive structure of the encoder, the $\KO(8,2)$ decoder decodes the message bits from top to bottom, similar in style to Dumer's decoding in \S\ref{sec:prob_formul}. More specifically, at any internal node of the tree we first decode the message bits along its left branch, which we utilize to decode that of the right branch and this procedure is carried out recursively till all the bits are recovered. At the leaves, we use the Soft-MAP decoder to decode the bits. 

Similar to the encoder $g_\theta$, an important aspect of our $\KO(8,2)$ decoder is a set of decoding neural networks $f_\phi = \{f_1, f_2,\ldots, f_{11}, f_{12} \}$. For each node $i$ in the tree, $f_{2i-1}:\reals^2 \to \reals$ corresponds to its left branch whereas $f_{2i}:\reals^4 \to \reals$ corresponds to the right branch. The pair of decoding neural networks $(f_{2i-1}, f_{2i})$ can be viewed as matching decoders for the corresponding encoding network $g_i$: While $g_i$ encodes the left and right codewords arriving at this node, the outputs of $f_{2i-1}$ and $f_{2i}$ represent appropriate Euclidean feature vectors for decoding them. Further, $f_{2i-1}$ and $f_{2i}$ can also be viewed as a generalization of Dumer's decoding to nonlinear real codewords: $f_{2i-1}$ generalizes the $\mathrm{LSE}$ function, while $f_{2i}$ extends the operation $\oplus_{\hat{\bv}}$. More precisely, we represent $f_{2i-1}(\by_1, \by_2)= \tilde{f}_{2i-1}(\by_1, \by_2) + \mathrm{LSE}(\by_1, \by_2)$ whereas $f_{2i}(\by_1, \by_2,\by_{\bv}, \hat{\bv} )= \tilde{f}_{2i}(\by_1, \by_2, \by_{\bv}, \hat{\bv}) + \by_1 + (-1)^{\hat{\bv}} \by_2$, where $(\by_1, \by_2)$ are appropriate feature vectors from the parent node, and $\by_{\bv}$ is the feature corresponding to the left-child $\bv$, and $\hat{\bv}$ is the decoded left-child codeword. We explain about these feature vectors in more detail below. Note that both the functions $\tilde{f}_{2i-1}$ and $ \tilde{f}_{2i}$ are also applied coordinate-wise.

{\bf Decoding.} At the decoder suppose we receive a noisy codeword $\by \in \reals^{256}$ at the root upon transmission of the actual codeword $\bx \in \reals^{256}$ along the channel. The first step is to obtain the LLR feature for the left $\RM(7,1)$ codeword: we obtain this via the left neural network $f_1$, i.e. $\by_v = f_1(\by_{1:128}, \by_{129:256}) \in \reals^{128}$. Subsequently, the Soft-MAP decoder transforms this feature into an LLR vector for the message bits, \ie $\bL_{(7,1)}= \text{Soft-MAP}(f_1(\by_{1:128}, \by_{129:256}))$. Note that the message bits $\bma_{(7,1)}$ can be hard decoded directly from the sign of $\bL_{(7,1)}$. Instead here we use their soft version via the sigmoid function $\sigma(\cdot)$, i.e., $\hat{\bma}_{(7,1)}= \sigma(\bL_{(7,1)})$. Thus we obtain the corresponding $\RM(7,1)$ codeword $\hat{\bv}$ by encoding the message $\hat{\bma}_{(7,1)}$ via an $\RM(7,1)$ encoder. The next step is to obtain the feature vector for the right child. This is done using the right decoder $f_2$, \ie $\by_{\bu} = f_2(\by_{1:128}, \by_{129:256}, \by_{\bv}, \hat{\bv})$. Utilizing this right feature $\by_{\bu}$ the decoding procedure is thus recursively carried out till we compute the LLRs for all the remaining message bits $\bma_{(6,1)}, \ldots, \bma_{(2,2)}$ at the leaves. Finally we obtain the full LLR vector $\bL = (\bL_{(7,1)}, \ldots, \bL_{(2,2)})$ corresponding to the message bits $\bma$. A simple sigmoid transformation, $\sigma(\bL)$, further yields the probability of each of these message bits being zero, \ie $\sigma(\bL)=\prob{\bma=\mathbf{0}}$.

Note that the decoding map $f_\phi:\by \mapsto \bL$ is fully differentiable with respect to $\phi$, which further ensures a differentiable loss for training the parameters $(\theta, \phi)$. 


\subsection{Training}

Recall that we have the following flow diagram from encoder till the decoder when we transmit the message bits $\bma$: $\bma  \xrightarrow{g_\theta} \bx \xrightarrow{Channel} \by \xrightarrow{f_\phi} \bL \xrightarrow{\sigma(\cdot)} \sigma(\bL)$. In view of this, we define an end-to-end differentiable cross entropy loss function to train the parameters $(\theta, \phi)$, \ie
\begin{align*}
    L(\theta, \phi ) = \sum_{j} m_j \log (1- \sigma(L_j)) + (1-m_j) \log \sigma(L_j).
\end{align*}
Finally we run Algorithm~\ref{alg:training} on the loss $L(\theta, \phi)$ to train the parameters $(\theta, \phi)$ via gradient descent.


\section{Soft-MAP decoder}
\label{sec:map}
As discussed earlier (see also Figure~\ref{Fig_RM82_Decoders}), Dumer's decoder for second-order RM codes $\RM(m,2)$ performs MAP decoding at the leaves while our \KO~decoder applies Soft-MAP decoding at the leaves. The leaves of both $\RM(m,2)$ and $\KO(m,2)$ codes are comprised of order-one RM codes and the $\RM(2,2)$ code. In this section, we first briefly state the MAP decoding rule over general binary-input memoryless channels and describe how the MAP rule can be obtained in a more efficient way, with complexity $\mathcal{O}(n\log n)$, for first-order RM codes. We then present the generic Soft-MAP decoding rule and its efficient version for first-order RM codes.

{\bf MAP decoding.} Given a length-$n$ channel LLR vector $\boldsymbol{l} \in \reals^n$ corresponding to the transmission of a given $(n,k)$ code, \ie code dimension is $k$ and block length is $n$, with codebook $\mathcal{C}$ over a general binary-input memoryless channel, the MAP decoder picks a codeword $\boldsymbol{c}^{*}$ according to the following rule \cite{abbe2020reedmuller}
\begin{align}\label{map}
\boldsymbol{c}^{*}=\operatorname*{argmax}_{\boldsymbol{c}\in\mathcal{C}}~~ \langle\boldsymbol{l},1-2{\boldsymbol{c}}\rangle,
\end{align}
where $\langle \cdot,\cdot\rangle$ denotes the inner-product of two vectors. Obviously, the MAP decoder needs to search over all $2^k$ codewords while each time computing the inner-product of two length-$n$ vectors. Therefore, the MAP decoder has a complexity of $\mathcal{O}(n2^k)$. Thus the MAP decoder can be easily applied to decode small codebooks like an $\RM(2,2)$ code, that has blocklength $n=4$ and a dimension $k=4$, with complexity $\mathcal{O}(1)$. On the other hand, a naive implementation of the MAP rule for $\RM(m,1)$ codes, that have $2^k=2^{m+1}=2n$ codewords, requires $\mathcal{O}(n^2)$ complexity. However, utilizing the special structure of order-1 RM codes, one can apply the fast Hadamard transform (FHT) to implement their MAP decoding in a more efficient way, i.e., with complexity $\mathcal{O}(n\log n)$. The idea behind the FHT implementation is that the standard $n\times n$ Hadamard matrix $\mathbf{H}$ contains half of the the $2n$ codewords of an $\RM(m,1)$ code (in $\pm 1$), and the other half are just $-\mathbf{H}$. Therefore, FHT of the vector $\boldsymbol{l}$, denoted by $\boldsymbol{l}_{\rm WH}$, lists half of the $2n$ inner-products in \eqref{map}, and the other half are obtained the as $-\boldsymbol{l}_{\rm WH}$. Therefore, the FHT version of the MAP decoder for first-order RM codes can be obtained as
\begin{align}\label{map_FHT}
\boldsymbol{c}^{*}=(1-{\rm sign}(\boldsymbol{l}_{\rm WH}(i^*))\boldsymbol{h}_{i^*})/2~~~~\text{s.t.}~~~~~i^*=\operatorname*{argmax}_{i\in[n]}~~ |\boldsymbol{l}_{\rm WH}(i)|,
\end{align}
where $\boldsymbol{l}_{\rm WH}(i)$ is the $i$-th element of the vector $\boldsymbol{l}_{\rm WH}$, and $\boldsymbol{h}_{i}$ is the $i$-th row of the matrix $\mathbf{H}$. Given that $\boldsymbol{l}_{\rm WH}$ can be efficiently computed with $\mathcal{O}(n \log n)$ complexity, the FHT version of the MAP decoder for the first-order RM codes, described in \eqref{map_FHT}, has a complexity of $\mathcal{O}(n \log n)$.

{\bf Soft-MAP.} Note that the MAP decoder and its FHT version involve ${\rm argmax}(\cdot)$ operation which is not differentiable. In order to overcome this issue, we obtain the soft-decision version of the MAP decoder, referred to as Soft-MAP decoder, to come up with differentiable decoding at the leaves \cite{RM_subcode}. The Soft-MAP decoder obtains the soft LLRs instead of hard decoding of the codes at the leaves. Particularly, consider an AWGN channel model as $\boldsymbol{y}=\boldsymbol{s}+\boldsymbol{n}$, where $\boldsymbol{y}$ is the length-$n$ vector of the channel output, $\boldsymbol{s}:=1-2{\boldsymbol{c}}$, $\boldsymbol{c}\in \mathcal{C}$, and $\boldsymbol{n}$ is the vector of the Gaussian noise with mean zero and variance $\sigma^2$ per element. The LLR of the $i$-th information bit $u_i$ is then defined as
\begin{align}\label{llrinf1}
\boldsymbol{l}_{\rm inf}(i) := \ln\left(\frac{\Pr(u_i=0|\boldsymbol{y})}{\Pr(u_i=1|\boldsymbol{y})}\right).
\end{align}
By applying the Bayes' rule, the assumption of  $\Pr(u_i=0)=\Pr(u_i=1)$, the law of total probability, and the distribution of the Gaussian noise, we can write \eqref{llrinf1} as
\begin{align}\label{llrinf2}
\boldsymbol{l}_{\rm inf}(i) =\ln\left(\frac{\sum_{\boldsymbol{s}\in\mathcal{C}_i^0}\exp\left(-||\boldsymbol{y}-\boldsymbol{s}||_2^2/\sigma^2\right)}{\sum_{\boldsymbol{s}\in\mathcal{C}_i^1}\exp\left(-||\boldsymbol{y}-\boldsymbol{s}||_2^2/\sigma^2\right)}\right).
\end{align}
We can also apply the max-log approximation to approximate \eqref{llrinf2} as follows.
\begin{align}\label{llrinf3}
\boldsymbol{l}_{\rm inf}(i) \approx 
\frac{1}{\sigma^2}\operatorname*{min}_{\boldsymbol{c}\in\mathcal{C}_i^1}||\boldsymbol{y}-\boldsymbol{s}||_2^2-\frac{1}{\sigma^2} \operatorname*{min}_{\boldsymbol{c}\in\mathcal{C}_i^0}||\boldsymbol{y}-\boldsymbol{s}||_2^2,
\end{align}
where $\mathcal{C}_i^0$ and $\mathcal{C}_i^1$ denote the subsets of codewords that have the $i$-th information bit $u_i$ equal to zero and one, respectively. Finally, given that the length-$n$ LLR vector of the cahhnel output can be obtained as $\boldsymbol{l}:=2\boldsymbol{y}/\sigma^2$ for the AWGN channels, and assuming that all the codewords $\boldsymbol{s}$'s have the same norm, we obtain a more useful version of the Soft-MAP rule for approximating the LLRs of the information bits as
\begin{align}\label{llrinf4}
\boldsymbol{l}_{\rm inf}(i) \approx \operatorname*{max}_{\boldsymbol{c}\in\mathcal{C}_i^0}~\langle \boldsymbol{l}, 1-2{\boldsymbol{c}}\rangle~-~ \operatorname*{max}_{\boldsymbol{c}\in\mathcal{C}_i^1}~\langle \boldsymbol{l}, 1-2{\boldsymbol{c}}\rangle.
\end{align}
It is worth mentioning at the end that, similar to the MAP rule, one can compute all the $2^k$ inner products in $\mathcal{O}(n2^k)$ time complexity, and then obtain the soft LLRs by looking at appropriate indices. As a result, the complexity of the Soft-MAP decoding for decoding $\RM(m,1)$ and $\RM(2,2)$ codes is $\mathcal{O}(n^2)$ and $\mathcal{O}(1)$ respectively. However, one can apply an approach similar to \eqref{map_FHT} to obtain a more efficient version of the Soft-MAP decoder, with complexity $\mathcal{O}(n\log n)$, for decoding $\RM(m,1)$ codes.

\section{Experimental details}
\label{sec:exp_detail}
We provide our code at  \href{https://github.com/deepcomm/KOcodes}{https://github.com/deepcomm/KOcodes}.


\subsection{Training algorithm}
\begin{algorithm2e}[ht]
   \caption{Training algorithm for \KO(8,2)} 
   \label{alg:training} 
   	\DontPrintSemicolon 
	\KwIn{number of epochs $T$, number of encoder training steps $T_{\rm enc}$, number of decoder training steps $T_{\rm dec}$, encoder training SNR ${\rm SNR}_{\rm enc}$, decoder training SNR ${\rm SNR}_{\rm dec}$, learning rate for encoder ${\rm lr}_{\rm enc}$, learning rate for decoder ${\rm lr}_{\rm dec}$}
	{\bf Initialize $(\theta, \phi)$}\;
	\For{$T$ \rm{steps}}{
	    \For{$T_{\rm dec}$ \rm{steps}}{
	        Generate a minibatch of random message bits $\bma$\;
	        Simulate AWGN channel with ${\rm SNR}_{\rm dec}$\;
            Fix $\theta$,  update  $\phi$ by minimizing $L(\theta, \phi)$  using Adam with learning rate  ${\rm lr}_{\rm dec}$\;
        }
        \For{$T_{\rm enc}$ \rm{steps}}{
            Generate a minibatch of random message bits $\bma$\;
            Simulate AWGN channel with ${\rm SNR}_{\rm enc}$\;
            Fix $\phi$, update  $\theta$ by minimizing $L(\theta, \phi)$    using Adam with learning rate  ${\rm lr}_{\rm enc}$\;
        }
  
    }
    \KwOut{$(\theta, \phi)$ } 

\end{algorithm2e} 
\subsection{Hyper-parameter choices for \KO(8,2)}
We choose batch size $B=50000$, encoder training SNR ${\rm SNR}_{\rm enc}=-3 dB$, decoder trainint SNR ${\rm SNR}_{\rm dec}=-5 dB$, number of epochs $T=2000$, number of encoder training steps $T_{\rm enc}=50$, number of decoder training steps $T_{\rm dec}=500$. For Adam optimizer, we choose learning rate for encoder ${\rm lr}_{\rm enc}=10^{-5}$ and for decoder ${\rm lr}_{\rm dec}=10^{-4}$.

\subsection{Neural network architecture of \KO(8,2)}
\label{sec:architecture}
\subsubsection{Initialization}
We design our (encoder, decoder) neural networks to generalize and build upon the classical (Plotkin map, Dumer's decoder). In particular, as discussed in Section~\ref{sec:enc82}, we parameterize the {\KO} encoder $g_\theta$, as $g_i(\bu, \bv) = (\bu, \tilde{g}_i(\bu, \bv) + \bu \oplus \bv )$, where $\tilde{g}:\reals^2 \to \reals$ is a fully connected neural network, which we delineate in Section~\ref{nn_table:g}. Similarly, for {\KO} decoder, we parametrize it as $f_{2i-1}(\by_1, \by_2)= \tilde{f}_{2i-1}(\by_1, \by_2) + \mathrm{LSE}(\by_1, \by_2)$ and $f_{2i}(\by_1, \by_2, \by_{\bv}, \hat{\bv} )= \tilde{f}_{2i}(\by_1, \by_2, \by_{\bv}, \hat{\bv}) + \by_1 + (-1)^{\hat{\bv}} \by_2$, where $\tilde{f}_{2i-1}:\reals^2 \to \reals$ and $\tilde{f}_{2i}:\reals^4 \to \reals$ are also fully connected neural networks whose architectures are described in Section~\ref{nn_table:f2} and Section~\ref{nn_table:f1}. If $\tilde{f}\approx 0$ and $\tilde{g}\approx 0$, we are able to thus recover the standard $\RM(8,2)$ encoder and its corresponding Dumer decoder. By initializing all the weight parameters $(\theta, \phi)$ sampling from ${\cal N}(0, 0.02^2)$, we are able to approximately recover the performance $\RM(8,2)$ at the beginning of the training which acts as a good initialization for our algorithm.

\subsubsection{Architecture of $\tilde{g}_{i}$} 
\label{nn_table:g}
\begin{itemize}
\itemsep0em
    \item Dense(units=$2\times32$)
    \item SeLU()
    \item Dense(units=$32\times32$)
    \item SeLU()
    \item Dense(units=$32\times32$)
    \item SeLU()
    \item Dense(units=$32\times1$)
\end{itemize}

\subsubsection{Architecture of $\tilde{f}_{2i}$}
\label{nn_table:f1}

\begin{itemize}
\itemsep0em
    \item Dense(units=$4\times32$)
    \item SeLU()
    \item Dense(units=$32\times32$)
    \item SeLU()
    \item Dense(units=$32\times32$)
    \item SeLU()
    \item Dense(units=$32\times1$)
\end{itemize}

\subsubsection{Architecture of $\tilde{f}_{2i-1}$}
\label{nn_table:f2}

\begin{itemize}
\itemsep0em
    \item Dense(units=$2\times32$)
    \item SeLU()
    \item Dense(units=$32\times32$)
    \item SeLU()
    \item Dense(units=$32\times32$)
    \item SeLU()
    \item Dense(units=$32\times1$)
\end{itemize}

\section{Results for Order-$1$ codes}
\label{sec:order1}

\begin{figure}[H] 
    \centering

    \begin{minipage}[c]{0.4\linewidth}
    \includegraphics[width=1\textwidth]{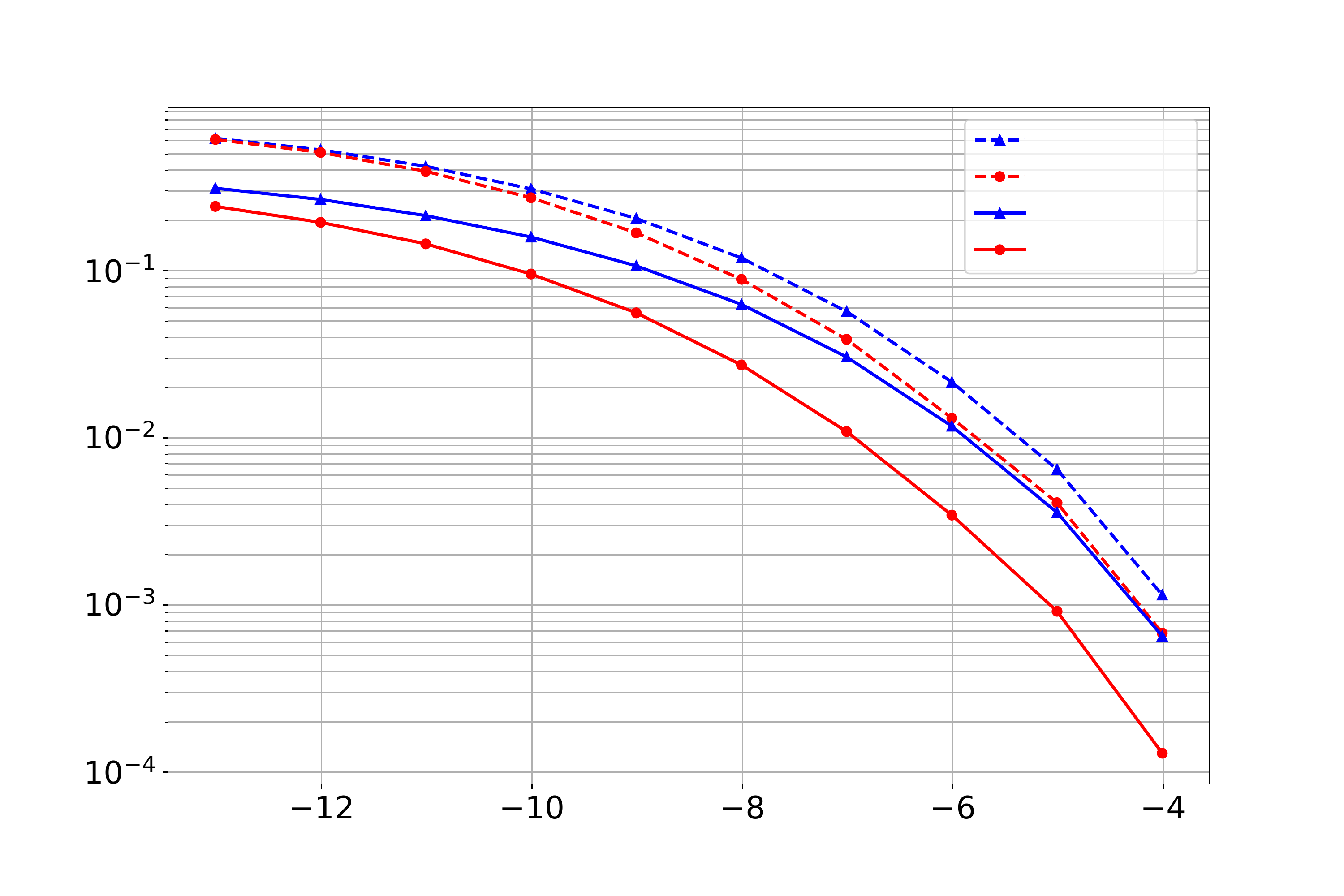}\put(-42,108){{\fontsize{5}{6}\selectfont RM BLER}}
    \put(-42,103){{\fontsize{5}{6}\selectfont\KO~BLER}}
    \put(-42,98){{\fontsize{5}{6}\selectfont RM BER}}
    \put(-42,93){{\fontsize{5}{6}\selectfont \KO~BER}}
    \put(-153,0){\footnotesize Signal-to-noise ratio (SNR) [dB]}
    \put(-192,63){\rotatebox[origin=t]{90}{\footnotesize Error rate}}
    \end{minipage}
    \begin{minipage}[c]{0.4\linewidth}
    \includegraphics[width=1\textwidth]{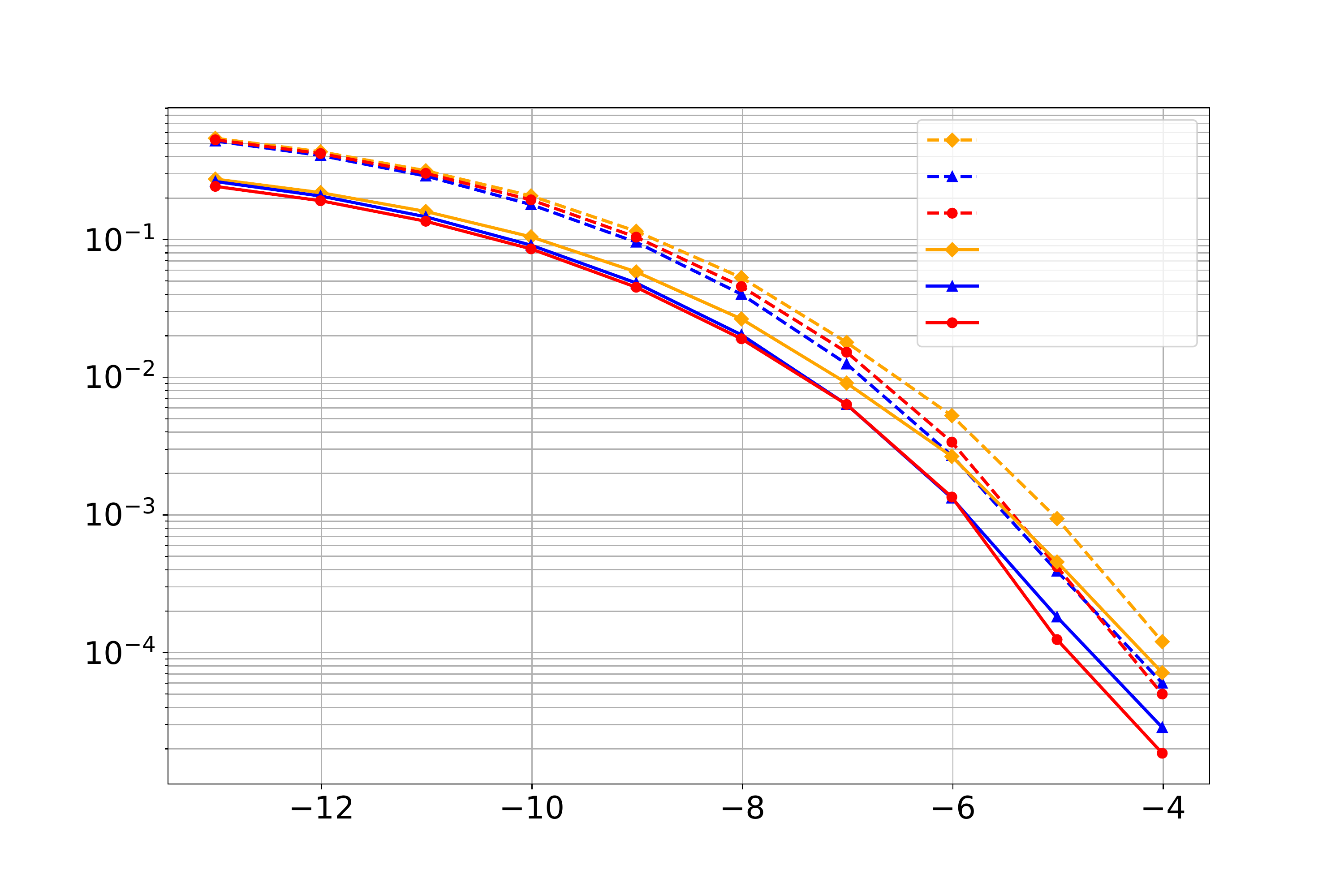}  \put(-52,108){{\fontsize{5}{6}\selectfont Gaussian BLER}}
    \put(-52,103){{\fontsize{5}{6}\selectfont RM~BLER}}
    \put(-52,98){{\fontsize{5}{6}\selectfont \KO~BLER}}
    \put(-52,93){{\fontsize{5}{6}\selectfont Gaussian~BER}}
    \put(-52,88){{\fontsize{5}{6}\selectfont RM~BER}}
    \put(-52,82){{\fontsize{5}{6}\selectfont \KO~BER}}
    \put(-153,0){\footnotesize Signal-to-noise ratio (SNR) [dB]}
    \put(-192,63){\rotatebox[origin=t]{90}{\footnotesize Error rate}} 
    \end{minipage}
            \caption{$\KO(6,1)$ code. {\bf Left}: $\KO(6,1)$ code achieves significant gain over $\RM(6,1)$ code (with Dumer) when trained on AWGN channel. {\bf Right}: Under the optimal MAP decoding, $\KO(6,1)$ and $\RM(6,1)$ codes achieve the same performance. Error rates for a random Gaussian codebook are also plotted as a baseline. }
            \label{fig:order1}
\end{figure}
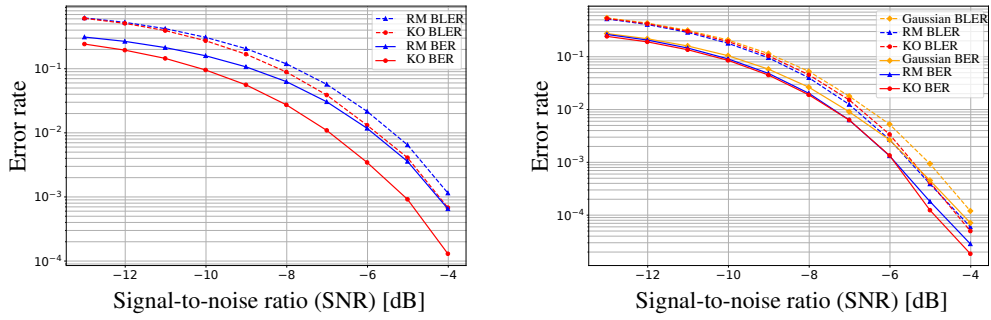

Here we focus on first order $\KO(m,1)$ codes, and in particular $\KO(6,1)$ code that has code dimension $ k= $7 and blocklength $n=64$. The training of the (encoder, decoder) pair $(g_\theta, f_\phi)$ for $\KO(6,1)$is almost identical to that of the second order $\RM(8,2)$ described in \S\ref{sec:neural_ko}. The only difference is that we now use the Plotkin tree structure of the corresponding $\RM(6,1)$ code. In addition, we also train our neural encoder $g_\theta$ together with the differentiable MAP decoder, \ie the Soft-MAP, to compare its performance to that of the RM codes. Figure~\ref{fig:order1} illustrates these results. 

The left panel of Figure~\ref{fig:order1} highlights that $\KO(6,1)$ obtains significant gain over $\RM(6,1)$ code (with Dumer decoder) when both the neural encoder and decoder are trained jointly. On the other hand, in the right panel, we notice that we match the performance of that of the $\RM(6,1)$ code (with the MAP decoder) when we just train the encoder $g_\theta$ (with the MAP decoder). In other words, under the optimal MAP decoding, $\KO(6,1)$ and $\RM(6,1)$ codes behave the same. Note that the only caveat for $\KO(6,1)$ in the second setting is that its MAP decoding complexity is $O(n^2)$ while that of the RM is $O(n\log n)$.

\end{document}